\documentclass[twocolumn,showpacs,amsmath,amssymb,prl,superscriptaddress,floatfix,aps]{revtex4-1}

\usepackage[utf8]{inputenc}
\usepackage[english]{babel}
\usepackage{graphicx}
\usepackage{dcolumn}
\usepackage{bm}
\usepackage{hyperref}

\usepackage{color}

\usepackage{amssymb}
\usepackage{amsmath}
\usepackage{latexsym}
\usepackage{amsthm}
\usepackage{latexsym}
\usepackage{newlfont}
\usepackage{textcomp}
\usepackage{amsfonts}
\usepackage{verbatim}

\begin{document}

\title{Cluster Luttinger Liquids of Rydberg-dressed Atoms in Optical Lattices}

\author{Marco Mattioli}
\affiliation{Institute for Quantum Optics and Quantum Information, Austrian Academy of Sciences, 6020 Innsbruck, Austria}
\affiliation{Institute for Theoretical Physics, University of Innsbruck, 6020 Innsbruck, Austria}

\author{Marcello Dalmonte}
\affiliation{Institute for Quantum Optics and Quantum Information, Austrian Academy of Sciences, 6020 Innsbruck, Austria}
\affiliation{Institute for Theoretical Physics, University of Innsbruck, 6020 Innsbruck, Austria}
\author{Wolfgang Lechner}
\affiliation{Institute for Quantum Optics and Quantum Information, Austrian Academy of Sciences, 6020 Innsbruck, Austria}
\affiliation{Institute for Theoretical Physics, University of Innsbruck, 6020 Innsbruck, Austria}
\author{Guido Pupillo}
\affiliation{ISIS (UMR 7006) and IPCMS (UMR 7504), University of Strasbourg and CNRS, Strasbourg, France}

\begin{abstract}
We investigate the zero-temperature phases of bosonic and fermionic gases confined to one dimension and interacting via a class of finite-range soft-shoulder potentials (i.e. soft-core potentials with an additional hard-core onsite interaction). Using a combination of analytical and numerical methods, we demonstrate the stabilization of critical quantum liquids with qualitatively new features with respect to the Tomonaga-Luttinger liquid paradigm. These features result from frustration and cluster formation in the corresponding classical ground-state. Characteristic signatures of these liquids are accessible in state-of-the-art experimental setups with Rydberg-dressed ground-state atoms trapped in optical lattices.
\end{abstract}

\pacs{71.10.Pm,05.30.Jp,32.80.Ee}

\maketitle
Frustration is responsible for a variety of phenomena in many-body physics, such as the emergence of gauge symmetries~\cite{magnet_book}, macroscopic ground-state degeneracies in strongly-correlated quantum spin and fermionic models~\cite{lee2006}, and glasses \cite{BINDERBOOK}. For particles with purely repulsive soft-core potentials, frustration can result from the self-assembly of conglomerated clusters~\cite{LIKOS2001,MLADEK2006}. This unusual phenomenon may open an experimental route towards the realization of exotic quantum phases in two and three dimensions, such as, e.g., supersolidity~\cite{Pomeau1994,Svistunov2005,Boninsegni2012,Henkel2010,Cinti2010,Cinti2013,kunimi2012} using Rydberg-dressed atomic gases~\cite{Pupillo2010,revexp,exp,Gallagher,Santos2001,Saffman2010,Lahaye2009,Bloch2008,Baranov2012}.

In one dimension (1D), gapless quantum systems seem to invariably fall within the universality class of Tomonaga-Luttinger (TL) liquids~\cite{haldane1981,GIAMARCHI2003,gogolin_book,cazalilla2004,cazalilla2011}. 
There, according to Haldane's construction~\cite{haldane1981}, the low-energy behavior is described by a free bosonic theory~\cite{difrancesco_book}, where the asymptotic decay of correlation functions in the microscopic model can be predicted in terms of just a few effective parameters. 
The TL liquid picture holds for both spin and Hubbard-like models with short-range interactions; such universal behavior is relevant to condensed matter systems~\cite{testll} as diverse as carbon nanotubes~\cite{carbon}, edge states in quantum Hall materials~\cite{hall} and systems with power-law-decaying interactions realizable in cold gases of ions~\cite{Ion}, atoms~\cite{Atom} and molecules~\cite{Mol1, Mol2}.  In short, the TL theory plays a similar key role in 1D to that of  the Fermi liquid theory in higher dimensions. While very recently possible signatures of a breakdown of the TL theory have been reported in certain quasi-1D electronic models~\cite{Fisher2013}, it is a fundamental question whether the conventional TL paradigm persists for any kind of interactions in 1D.

\begin{figure}[ht]
\centerline{\includegraphics[width=0.85\columnwidth]{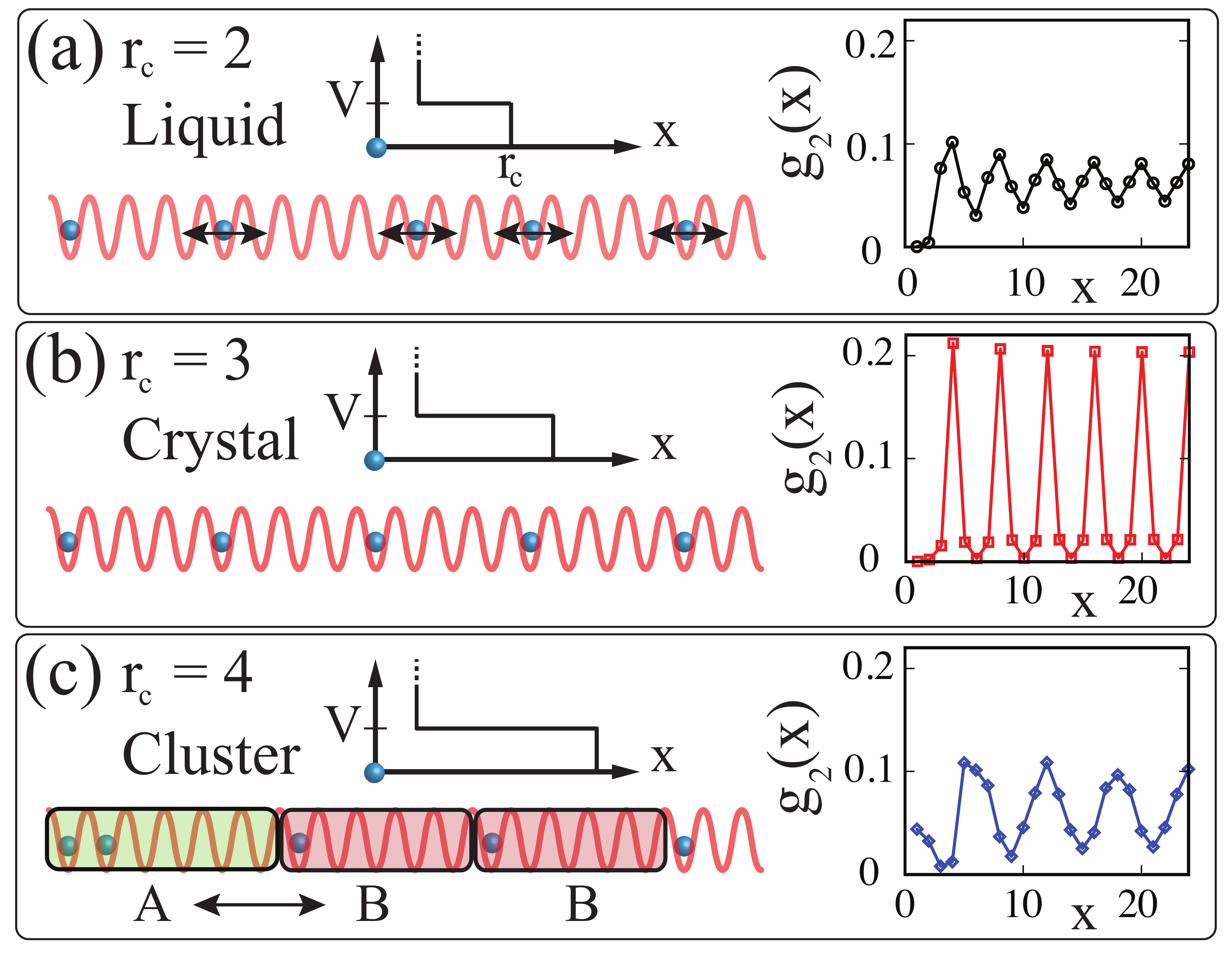}}
\caption{Sketches of the classical ground-states ($t=0$) of Eq.~\eqref{equ:hamiltonian} in the liquid (a), crystalline (b) and cluster (c) phases, for $r_c=2,3$ and $4$, respectively, and particle density $\overline{n}=1/4$. Particles are drawn in the lattice as blue spheres. Insets: numerical DMRG computations of density-density correlations $g_2(x)$ for $V/t=6$ and $L=48$ in the corresponding quantum phases. For $r_c=4$ [panel (c)], particles form two kinds of clusters associated to blocks A and B with two and one particles, respectively, followed by four empty sites.  In the quantum regime, this results in a cluster-type liquid with correlation functions different from a standard TL liquid.}
\label{fig:illustration}
\end{figure}

Here, we show that soft-shoulder potentials on 1D lattice systems lead to quantum liquid phases that do not fall into the conventional TL paradigm. Characteristic features of these anomalous {\it cluster Luttinger} (cL) liquids include a deformation of the critical surface in momentum space at odds with conventional TL physics, and are evident in correlation functions such as momentum distributions and structure factors. We trace back the origin of this anomalous behavior to cluster formation, which provides a natural source of frustration and leads to infinite ground-state degeneracies. We derive an effective low-energy quantum field theory based on a classical model of clusters, and confirm the departure from the TL picture by means of density-matrix-renormalization-group (DMRG)~\cite{White,Sch} numerical simulations.  We propose the observation of cL liquids in experiments with Rydberg-dressed ground-state alkali atoms loaded into optical lattices.

The relevant microscopic Hamiltonian for hard-core bosons or spinless fermions in 1D geometry reads
\begin{eqnarray}
\label{equ:hamiltonian}
H&=&-t\sum_i (b^\dagger_ib_{i+1}+\mathrm{c.c.})+
V\sum_{i}\sum_{\ell=1}^{r_c}n_in_{i+\ell}.
\end{eqnarray}
Here, $b^{\dagger}_i (b_i)$ are hard-core bosonic/fermionic creation (annihilation) operators at the site $i$, $n_i=b^{\dagger}_i b_i$, and $t$ is the tunnelling rate on a lattice of spacing $a$, which, from now on, we set as unit length of distances (the second sum only runs to the right to avoid double-counting). In the following, $\rho=\overline{n}$ ($\rho =1- \overline{n}$) for $\overline{n} \leq 1/2$ ($\overline{n} > 1/2$), with $\overline{n}$ the particle density.
The \emph{soft-shoulder} potential in Eq.~\eqref{equ:hamiltonian} can be engineered in clouds of cold Rydberg atoms,
where both the strength $V$ and the range $r_c$ of the soft-core contribution can be tuned by weakly-admixing the Rydberg level to the ground-state, whereas the additional onsite hard-core constraint is enforced using, e.g., Feshbach resonances. In the following, we focus on the bosonic case, discussing the fermionic one at the end of the manuscript.

Equation~\eqref{equ:hamiltonian} has a rich phase diagram, resulting from the combination of the following features of soft-core interactions: (i) their Fourier transform has negative components, which in classical systems are associated with free-space cluster formation \cite{LIKOS2001}; (ii) the competition of $r_c$ with the critical length $r^{\star}=1/\rho -1$ leads to frustration for $r_c>r^{\star}$. These effects result in the formation of highly degenerate cluster-type ground-states in the classical limit ($t=0$), as well as in novel phases in the quantum regime ($t>0$). Note that all correlation functions are here particle-hole symmetric, such that the phases for densities $\overline{n}$ and $1-\overline{n}$ are identical.

Fig.~\ref{fig:illustration} shows sketches of the possible ground-states for $\overline{n}=1/4$ ($r^\star=3$) and $t=0$. These phases are: a liquid $(r_c<r^{\star})$, a crystal $(r_c=r^{\star})$, and a cluster state $(r_c>r^{\star})$. The classical  liquid [Fig.~\ref{fig:illustration}(a)] is given by the summation of all particle configurations with the constraint that inter-particle distances are larger than $r_c$, while the crystal has a particle every $r_c+1$ lattice site  [Fig.~\ref{fig:illustration}(b)]. The ground-state in the cluster regime is instead highly degenerate. Its structure can be computed analytically using the following procedure: (i) for $r_c>r^{\star}$ particles and holes are grouped into blocks of type A and B consisting of two and one particles, respectively, followed by a number of $r_c$ empty sites [in Fig.~\ref{fig:illustration}(c), the example of $r_c = 4$ is shown]. Other block sizes are energetically unfavored. The total number of blocks $M$ for a system of length $L$ is $M=L(1-\rho)/r_c$. (ii) Any exchange of two blocks does not change the energy of the system. The classical ground-state then consists of all permutations of blocks A and B (e.g. [AABAAB], [ABAAAB], $...$) and the associated ground-state degeneracy $d = M!/[(M/3)! (2M/3)!]$ grows exponentially with $M$.

\begin{figure}[t]
\centerline{\includegraphics[width=8.0cm]{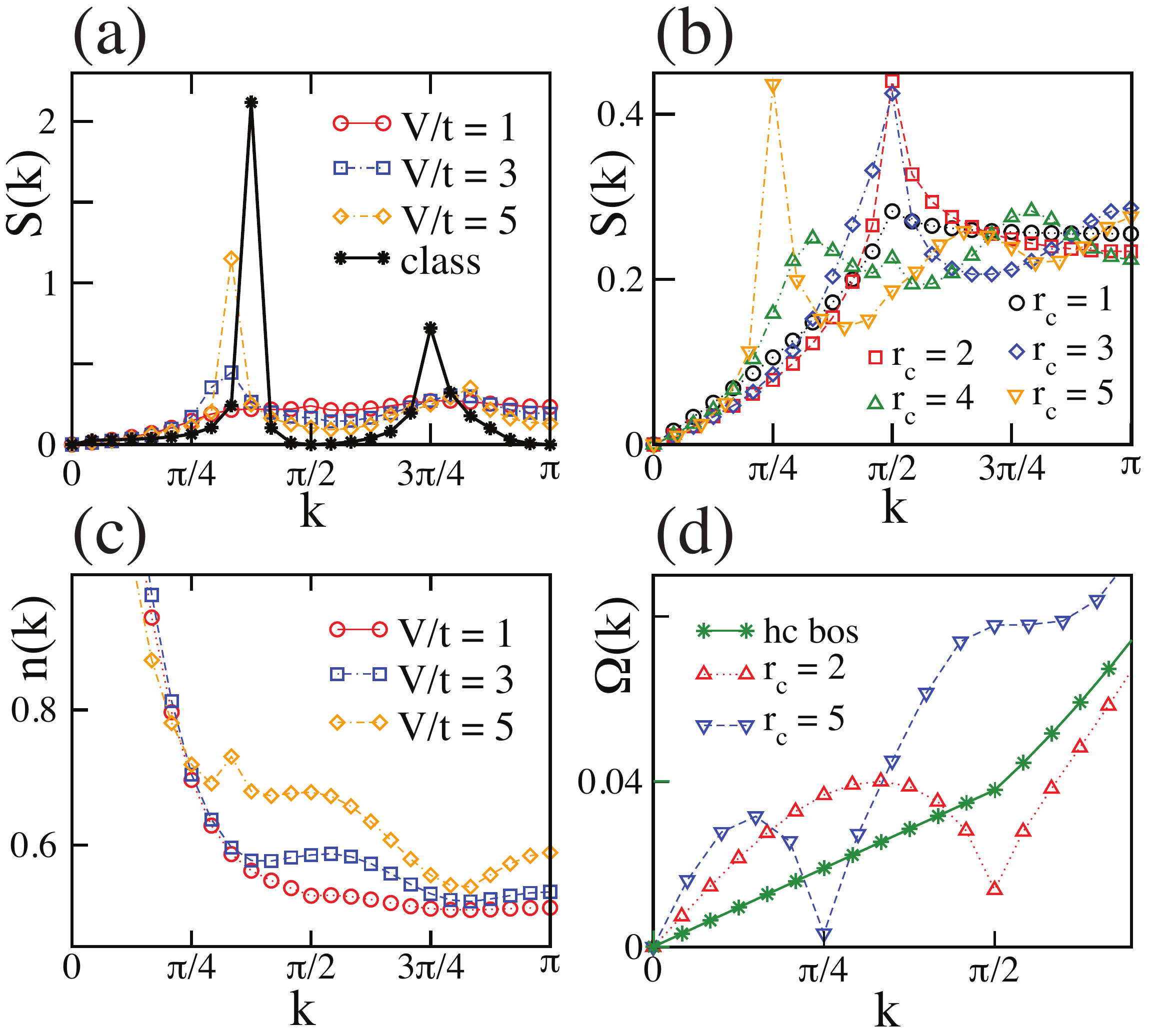}}
\caption{(a) Static structure factor $S(k)$ at fixed $r_c=4$ and several $V/t$ for a chain of length $L = 48$. Black solid line: classical case ($t=0$). (b)  $S(k)$ for different values of $r_c$ and $V/t=1.5$.  (c) Momentum distribution for the same parameters in (a). (d) Dispersion relation at $V/t=3$: for $r_c=2$ ($r_c=5$) a density roton minimum (cluster-like roton minimum) appears at the corresponding peak in $S(k)$. }\label{fig:results}
\end{figure}

The formation of the block structure above is well captured by the static structure factor $S(k)=\sum_{\ell,j}e^{ik(\ell-j)}g_2(\ell-j)/L$, with $g_2(\ell-j)=\langle n_\ell n_j\rangle - \overline{n}^2$ the density-density correlation function. This is shown in Fig.~\ref{fig:results}(a) for $\overline{n}=3/4$, $r_c=4 > r^{\star}$ and $t=0$, where $S(k)$ peaks at a $k$-value $k_c < \pi/2$ (solid black line). Numerical DMRG results in the same panel show that this peak persists for $t\simeq V$, where the ground-state is a strongly-interacting quantum liquid. This signals a departure from the TL picture, where $k_c = \pi/2$ for any shape of interactions. Key questions are: (i) what is the underlying mechanism and (ii) what is the effective theory  connecting the classical and quantum regimes.

\paragraph{Low-energy theory-} For bosonic and fermionic gapless systems, the low-energy physics is usually well captured by the universality class of the TL liquid. In Haldane's formulation, the operators $b_j$ are phenomenologically described by long-wavelength density and phase fluctuation fields, $\varphi(x)$ and $\vartheta(x)$ respectively,
satisfying $[\vartheta(y),\nabla\varphi(x)] = i\pi\delta(x - y)$,
and the density operator reads
\begin{equation}\label{psi_bos}
n(x)=[\overline{n}-\nabla\varphi(x)'/\pi]\sum_{p\in\mathbb{Z}} e^{2ip(\pi \overline{n}  x -\varphi'(x))},
\end{equation}
where $\varphi'(x)=\pi \overline{n} x-\varphi(x)/2$. The discreteness of particle density is captured
by the last factor in Eq.~\eqref{psi_bos}, which determines the behavior
of correlation functions beyond the hydrodynamic approximation, as, for example, $g_2(x) = \overline{n}^2+K/(2\pi x^2)+\alpha \cos(2\pi \overline{n} x)/x^{2K}$. Here, $\alpha$ is a non-universal constant and $K$ is
the Luttinger parameter, which embodies the effect of interactions in the low-energy theory.  The form of $g_2(x)$ implies peaks (if any) in $S(k)$ at momenta commensurate with density excitations, independently of the form and strength of interactions: these features characterize the regime $r_c \leq r^\star$ [see Fig. \ref{fig:results}(b)]. For $r_c > r^{\star}$, however, the peak shifts to lower values $k_c<\pi/2$;
this feature, as well as others discussed below, is incompatible with a TL liquid~\cite{note_sg}.

In the following, we show that a constrained effective theory incorporating cluster-type effects is able to capture this phenomenology.
Our formalism is based on two assumptions: {\it i)} the low-energy dynamics takes place in the sub-manifold defined by the cluster states, where the number of clusters is fixed by the classical ground-state structure; {\it ii)} the internal cluster degree of freedom is frozen, that is, one can treat clusters of type A and B on equal footing. In Ref.~\cite{supmat} we demonstrate that both assumptions are well justified in the limit $V\gg t$.

\begin{figure}[bt]
\centerline{\includegraphics[width=0.44\textwidth]{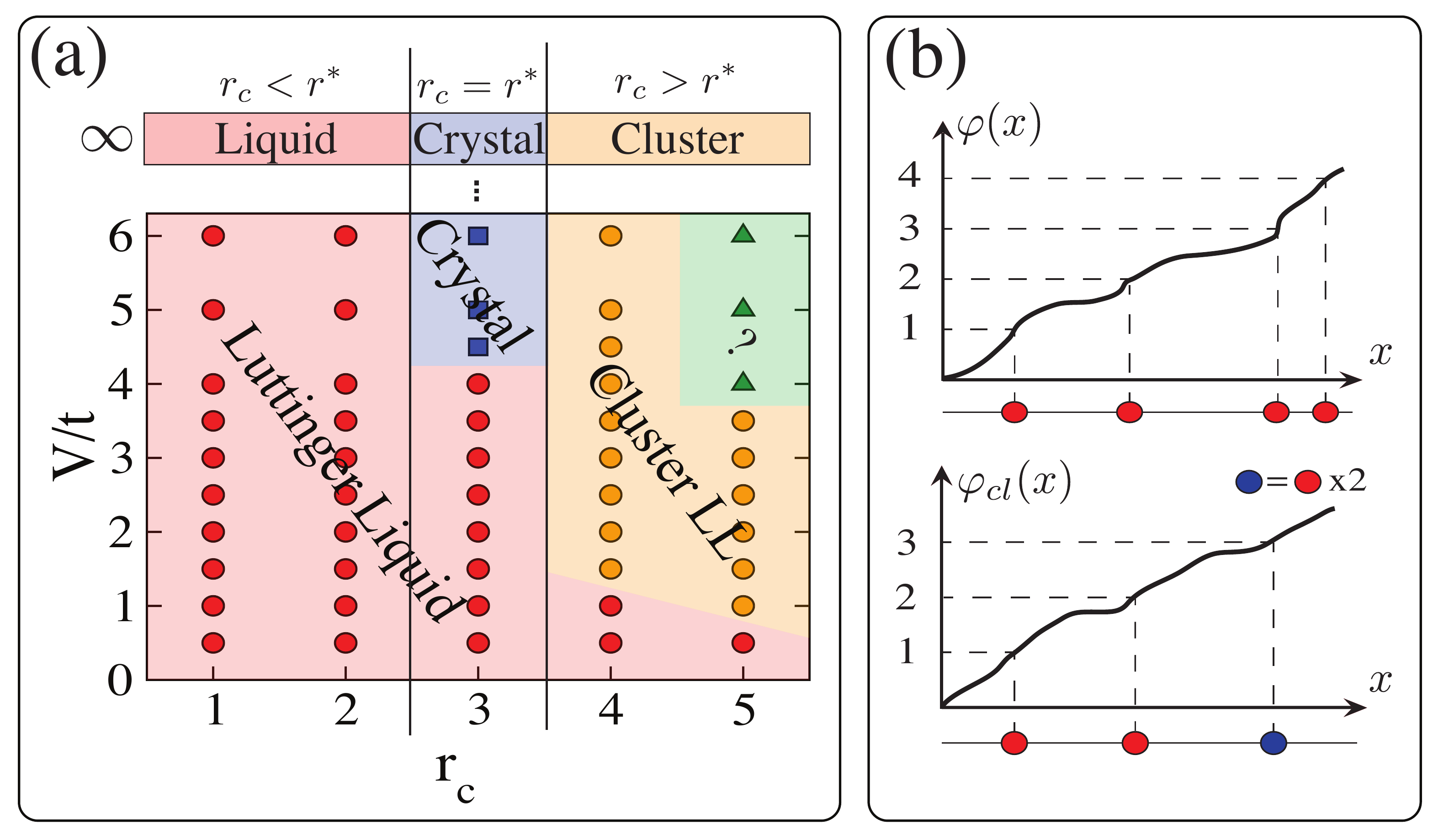}}
\caption{(a) Numerical zero-temperature phase diagram of Eq.~\eqref{equ:hamiltonian} for density $\overline{n}=3/4$. The following phases are indicated: TL liquid (red circles), crystal (blue squares), cL liquid (yellow circles). Green triangles define a phase which shows an anomalous
entanglement entropy growth (see main text). (b) Configurations allowed for the quantum field $\varphi(x)$: in the TL scenario (top), $\varphi(x)$ takes integer values at the position of each particle (red circles). In the cL liquid (bottom), $\varphi_{cl}$ is related to the cluster density, and not the particle density. This points to a different nature of collective excitations.}
\label{fig_varphi}
\end{figure}

In the low-density limit $r_c \overline{n}\gg1$ and $\overline{n}\ll 1$, we rewrite the density as $n(x)=\sum_{m=1}^Mf(x_m)\delta(x-x_m)$. Here $x_m$ indicates the position of the $m$-th cluster, whereas $f(x_m)\in \{1,2\}$ represents the $m$-th cluster size. We then introduce a
collective {\it cluster counting} field $\varphi_{cl}(x)$, so that the density operator
can be rewritten as $n(x)=\nabla\varphi_{cl}(x)\sum_{m=1}^Mf(x_m)\delta[\varphi_{cl}(x)-\pi m]$.
The field $\varphi_{cl}(x)$ differs sharply from its counterpart
$\varphi(x)$ in the standard TL theory, in that it takes integer values at the position of clusters and not of individual particles, as exemplified in Fig.~\ref{fig_varphi}(b).
After introducing $\varphi_{cl}'(x)=(2\pi \overline{n} \sigma x-\varphi_{cl}(x)/2)$,
with $\sigma=M/N$ the cluster density ($N$ being the total number of particles),
the corresponding functional dependence of $n(x)$ on $\varphi_{cl}'(x)$ reads
\begin{equation}\label{rho_cLL}
n(x)=
\left[ \overline{n} - \frac{1}{\sigma\pi} \nabla \varphi_{cl}(x)' \right]\sum_{k=-\infty}^\infty a_ke^{2i k[\varphi_{cl}(x)'-\pi \overline{n}\sigma x]},
\end{equation}
with $a_k$ constants. The resulting low-energy field theory stemming from Eq.~\eqref{equ:hamiltonian} is then a bosonic theory
\begin{equation}\label{Hle}
\mathcal{H}=\frac{v_{cl}}{2}\int dx [(\sqrt{4\pi} \partial_x\vartheta_{cl}/R_{\varphi})^2 + (R_{\varphi} \partial_x\varphi_{cl}/\sqrt{4\pi})^2]
\end{equation}
corresponding to a $c=1$ conformal field theory (a compactified boson model~\cite{difrancesco_book}) with sound velocity $v_{cl}$ and compactification radius $R_{\varphi}$ for the field $\varphi_{cl}$, whose conjugate phase field is $\vartheta_{cl}$.
According to assumptions {\it i)} and {\it ii)} above, the physical interpretation of the low-energy theory is then that of a quantum liquid of clusters: the suppression of the internal degree of freedom~\cite{supmat} leaves a single quantum field $\varphi_{cl}$, representing collective cluster excitations acting on top of the zero-energy degenerate cluster manifold.
Equations~\eqref{rho_cLL} and~\eqref{Hle} imply that physical observables will differ from those of TL liquids, since the microscopic bosonic fields have a different functional dependence with respect to the collective field $\varphi_{cl}$,  compared to the one indicated in Eq.~\eqref{psi_bos}. For example, $g_2(x)$ now reads $g_2(x) \simeq \overline{n}^2 + \alpha_1 / x^2 + [\alpha_2 \cos(2\pi \overline{n} \sigma x)]/x^\gamma$, with $\alpha_1$, $\alpha_2$ and $\gamma$ constants. Here, unlike in TL liquids, the relation $\alpha_1 = \gamma/(4 \pi)$ is not satisfied, preventing any direct estimate of correlation function decay from conventional techniques~\cite{supmat}.
The position $k_c$ of peaks in $S(k)$ now depends on the interplay between density and range of interactions, and is predicted to be $k_c=2\overline{n} \pi \sigma = 2 \pi (1-\rho )/r_c$, with $k_c < \pi/2$. For $r_c=4$ and $\rho=1/4$, e.g., $k_c=3\pi/8$, in very good agreement with classical numerical results [see Fig.~\ref{fig:results}(a)]. 
Notable differences occur in a similar way to several other physical observables: in Ref.~\cite{supmat}, we provide a detailed comparison with the numerical analysis, discussing discrepancies between TL and cL liquids based on level spectroscopy methods and correlation function decay. We note that the procedure above allows one to retain high-momentum features due to interactions in the low-energy theory (here corresponding to cluster formation), and can be in principle applied to other problems where the classical ground-state presents infinite degeneracies. Below we focus on the regimes $r_c<r^{\star}$, $r_c=r^{\star}$, and $r_c>r^{\star}$ using the model in Eq.~\eqref{equ:hamiltonian} as well as a soft-shoulder potential with an additional Van der Waals long-range tail. In Ref.~\cite{supmat}, we further corroborate evidence of the generality of our results analyzing Eq.~\eqref{equ:hamiltonian} for several $\overline{n}$, each of them supporting different ground-state configurations in the corresponding classical limit.

\paragraph{Numerical results -} We investigate the phase diagram of Eq.~\eqref{equ:hamiltonian}
by performing numerical DMRG simulations for a given particle density $\overline{n}$ as a function of $V/t$ and $r_c$. 
The case of $\overline{n}=3/4$ is shown in Fig.~\ref{fig_varphi}(a). There, for sufficiently small interactions, the low-energy phase is a TL liquid, independently of $r_c$. This is clearly seen for, e.g., $r_c =2 < r^{\star}$, where the system is equivalent to an XXZ model with next-nearest-neighbor interactions and finite magnetization. For $r_c = r^{\star}$, instead, the TL liquid turns into an insulating Mott phase with a hole every four sites [blue squares, see also sketch in Fig.~\ref{fig:illustration}(b)],  via a quantum phase transition of the Berezinskii-Kosterlitz-Thouless (BKT) type around $V/t \simeq 4$~\cite{supmat}.
The most interesting behavior occurs for $r_c>r^{\star}$ and $V/t \gtrsim 1$, where the TL turns into a cL liquid. Below we analyze in detail this regime, and comment on the region with $V/t \gg 1$ and $r_c \gtrsim 5$ (green triangles) towards the end of the paper.

We characterize the phase diagram by focussing on:
{\it i)} $S(k)$;
{\it ii)} the asymptotic decay of $g_2(\ell -j)$ and of the one-body density matrix
$B(\ell - j)=\langle b_{\ell}^{\dagger}b_{j}\rangle$; {\it iii)} the excitation spectrum
$\Omega(k)=E_k/S(k)$ in the Feynman approximation, with $E_k = \frac{t}{L}\left[ 1 -\cos\left(\frac{2\pi k}{L} \right)\right] \langle 0 | \sum_{i}( b^{\dagger}_i b_{i+1} + h.c. )| 0 \rangle$. $\Omega(k)$ is an upper bound to the exact spectrum~\cite{FENYMAN}.

According to the discussion above, differences between TL and
cL liquids are evident in $S(k)$ and $g_2(x)$: in particular, in the TL phase, $S(k)$ exhibits a peak at $k_{c}=\pi/2$, commensurate with the particle density, while peaks commensurate
with the cluster structure emerge in the cL phase, the exact position
depending on $r_c$. Figs.~\ref{fig:results}(a) and (b) demonstrate the crossover from TL to cL liquid
increasing the strength of interactions $V/t$ at fixed $r_c > r^{\star}$ and as a function of $r_c$ at fixed $V/t$, respectively.
In Ref.~\cite{supmat}, we show that correlation functions such as $B(x)$ decay algebraically in the TL and cL phases, validating the assumption that both underlying field theories are $c=1$ conformal. However, a strong oscillatory behavior for  $B(x)$ in the cL phase leads to the emergence of new peaks in the momentum distribution $n(k) = \sum_{\ell,j} e^{ik(\ell-j)}B(\ell - j)/L$ [Fig.~\ref{fig:results}(c)]. This effect, which may be observed experimentally, is predicted by our low-energy field theory and is reminiscent of the deformation of the {\it Bose surface} observed in frustrated, two-dimensional bosonic models~\cite{rigol}. Finally, we note that established methods to extract $K$ fail in the cL phase by giving non-consistent estimates~\cite{supmat}, further strengthening the departure from the standard TL scenario.

The excitation spectrum is analyzed in Fig.~\ref{fig:results}(d), where $\Omega(k)$ is plotted as a function of $k$ for several values of $r_c$ with fixed $V/t=3$. We find that the finite-range interactions within the regime $r_c < r^{\star}$ induce a roton excitation at $k_c=\pi/2$, commensurate with density excitations, as expected. For $r_c > r^{\star}$, however, $\Omega(k)$ develops a new roton-type excitation at $k_c < \pi/2$, reflecting the structure of the classical cluster ground-state.
Finally, in the green region with $V/t \gg 1$ and $r_c \gtrsim 5$ of Fig.~\ref{fig_varphi}(a), we observe both (i) correlation functions consistent with a cL liquid and (ii) a significant increase of the entanglement entropy. This might point towards a critical phase with central charge $c>1$,
due to the emergence of internal degrees of freedom in the block structure. However, strong finite-size effects prevent here any conclusive statement on this regime, which we plan to investigate in a future work. The phase diagram of Fig.~\ref{fig_varphi}(a) holds for fermions as well, where the
TL phase displays dominant density-wave order; hints on possible physics beyond TL
for similar Fermi systems have been reported in Refs.~\cite{Zhuravlev2000,KNAP2012}.

\begin{figure}[t]
\includegraphics[width=8.5cm]{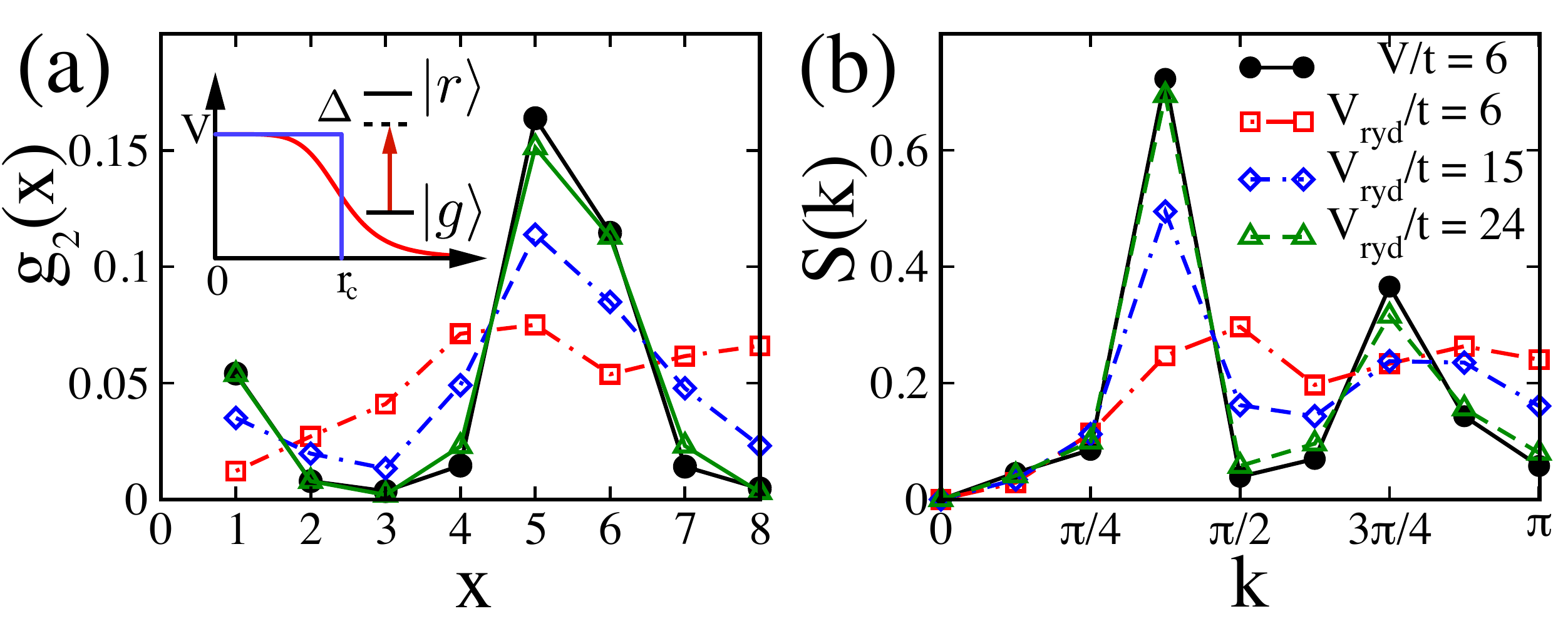}
\caption{(a) Inset: Laser-dressing scheme~\cite{supmat} and resulting Born-Oppenheimer potential $V_{\mathrm{ryd}}(i,j)$ for Rydberg-dressed atoms (red), compared to the potential in Eq.~\eqref{equ:hamiltonian} (blue). (a): $g_2(x)$ for several $V_{\mathrm{ryd}}/t$ and $r_c=4$ with $L=16$ [symbols as in (b)]. (b) $S(k)$ for several $V_{\mathrm{ryd}}/t$ in comparison to $V/t=6$ (black circles). The convergence of the observables for $V_{\mathrm{ryd}}/V \gtrsim 2$ can be understood from simple energy considerations~\cite{supmat}.}
\label{fig:k}
\end{figure}

Among all potentials which favor the stabilization of cL liquids, an interesting example is $V_{\mathrm{ryd}}(i,j) =  V_{\mathrm{ryd}}/\{1+[(i-j)a/r_c]^6\}$, realizable with cold Rydberg atoms~\cite{supmat}. Exact diagonalization results in Fig.~\ref{fig:k} show that, for $V_{\mathrm{ryd}}(i,j)$, features of $g_2(x)$ and $S(k)$ characteristic of cL liquids appear and converge to those of Eq.~\eqref{equ:hamiltonian}, provided that $V_{\mathrm{ryd}}/V \gtrsim 2$~\cite{supmat}. 

In conclusion, we have shown that soft-shoulder potentials support quantum liquid phases beyond the standard TL paradigm. The breakdown of TL theory has relevant theoretical and experimental consequences: e.g., it has been recently suggested as a possible descendant mechanism for {\it strange metal} phases~\cite{lee2006} of cuprate high-temperature superconductors. Here we have shown, in a completely different scenario of immediate interest for experiments with ultracold Rydberg atoms, that it can in fact emerge in a broad class of 1D quantum systems which support cluster formation in the classical limit. We hope that our work will provide new insights for unveiling other general mechanisms leading to anomalous TL physics.
Interesting extensions might include the search for exotic phases in 2D, such as frustration-induced Bose metals~\cite{block2011} and emergent gauge fields~\cite{magnet_book}.

\paragraph{Acknowledgements-} We thank M. Cheneau, F. Cinti, A. Gl\"atzle, C. Gross, A. L\"auchli, R. Nath, T. Pohl, P. Zoller for useful discussions, C. Laflamme and M. Rider for careful reading of the manuscript, and F. Ortolani for help with the
DMRG code. M.M. acknowledges the Marie Curie Initial Training Network COHERENCE for financial support. M.D. acknowledges support by the European Commission via the integrated project AQUTE, and
by the Austrian Science Fund FWF (SFB FOQUS F4015-N16). W.L. acknowledges support by the Austrian Science Fund through P 25454-N27 and by the Institut f\"ur Quanteninformation. G.P. acknowledges KITP in Santa Barbara for hospitality and support by the European Commission via ERC-St grant "ColdSIM" (grant agreement 307688), EOARD, and UdS via Labex NIE and IdEX.

\newpage

\onecolumngrid

\vspace{1.5cm}

\begin{center}
{\bf \large Supplementary Material for ``Cluster Luttinger Liquids of Rydberg-dressed Atoms in Optical Lattices''} \\

\vspace{0.6cm}

{Marco Mattioli$^{1,2}$, Marcello Dalmonte$^{1,2}$, Wolfgang Lechner$^{1,2}$, and Guido Pupillo$^3$ } \\
{$^1$ {\it Institute for Quantum Optics and Quantum Information, Austrian Academy of Sciences, 6020 Innsbruck, Austria}}\\
{$^2${\it Institute for Theoretical Physics, University of Innsbruck, 6020 Innsbruck, Austria}} \\
{$^3${\it ISIS (UMR 7006) and IPCMS (UMR 7504), University of Strasbourg and CNRS, Strasbourg, France}}\\
\end{center}

\vspace{0.6cm}

\twocolumngrid

\section{Experimental setup}
The phases of Eq.(1) of the main text should be realizable in state-of-the-art experimental setup of ultra-cold Rydberg-dressed atoms trapped in 1D optical lattices~\cite{Bloch2008}. We note that Rydberg-dressing in optical lattices has been very recently investigated in detail in Ref.~\cite{macri}. 

Here we consider, e.g., $^{85}$Rb atoms, where each ground-state  $\lvert g \rangle$ is off-resonantly coupled to an excited Rydberg state
$\lvert r \rangle$ via a two-photon laser with effective Rabi frequency
$\Omega$ and red detuning $\Delta$, where $|\Delta| \gg  \Omega$, [see inset of Fig.4(a), main text].
We choose $| r \rangle \equiv |43\,S_{1/2}\rangle$,
$\Omega/(2 \pi) =  2.3$ MHz and $\Delta/(2 \pi) =  20$ MHz, where we have set the Planck' s constant $h = 2 \pi$. Within these parameters, $V_{\mathrm{ryd}}/(2 \pi) = \Omega^4/(16 \pi \Delta^3) \simeq  440$ Hz. As already mentioned in the main text, the onsite hard-core constraint is enforced using external fields, such as Feshbach resonances. The resulting cut-off radius $r_c = (C_6/2\Delta)^{1/6} \simeq  1.98 \, \mu$m is considerably larger than usual lattice spacings $a \simeq 0.39\, \mu$m,
allowing one to access the cluster regime (here, $r_c/a = 5$).
We choose a lattice depth $V_0\gtrsim 13 E_R = 2\pi \cdot 50$ kHz, such that
the longitudinal trapping frequency $\omega_{\parallel}/(2 \pi) = E_R\sqrt{V_0/E_R} \simeq 13.9 \; \mathrm{kHz} > \max\{t,V_{\mathrm{ryd}}\}$  limits the dynamics to the lowest lattice band.
Here, $E_R=\pi^2 /(2ma^2)$ is the recoil energy of atoms with mass $m$, while  $t/(2 \pi)\lesssim 45$ Hz (see Ref.~\cite{Bloch2008}) is the hopping amplitude in the lattice. The strongly-interacting regime is easily accessible
since, within the chosen parameters, $V_{\mathrm{ryd}}/t \simeq 10$, ensuring the possibility to observe cL liquids for $r_c/a = 5$ (cfr. Fig.3(a) of the main text together with the requirement $V_{\mathrm{ryd}}/V \gtrsim 2$).
Transversal trapping frequencies $\omega_{\perp} \gg \omega_{\parallel}$ confine the system to a 1D geometry.
Finally, the small effective decay rate of the dressed ground-state results $\gamma_{\mathrm{eff}}/(2\pi) = (\Omega / 2\Delta)^{2} \gamma_r \simeq 1-10$ Hz ($\gamma_r$
being the bare decay rate of the Rydberg state $\lvert r \rangle$~\cite{beretov2009}). As the number
of particles needed to observe the distinguishing features of cL liquids is relatively 
small (in Fig.4 of the main text, 4 atoms in a system of 16 sites were shown to 
be sufficient to prove the emergence of anomalous correlations), such a small
decay rate should be sufficient to realize dominant Hamiltonian dynamics. 
Alternatively, one could employ dressing to p-states via a single-photon
transition, which allows for much larger Rabi frequencies and detunings, 
further reducing the influence of spontaneous emission and single-particle losses.

\section{Effective cluster model}

As explained in the main text, classically, the ground-state degeneracy for $r_c>r^{\star}$ can be explained introducing A-type building blocks (or "A-blocks") and B-type building blocks (or "B-blocks"). The classical ground-state is the starting point of our low-energy field theory, which is supplemented by two key assumptions: ({\it i}) all clusters are treated as single objects without internal correlations, and ({\it ii}) the mass difference between clusters can be neglected. These two assumptions allow us to describe the low-energy physics in terms of a single field. 

In the following, we discuss the physical interpretation of these approximations and their range of applicability. In Section~\ref{ssec:classical}, we show, from classical considerations, that clusters are stable due to the presence of large energy barriers, which prevent hopping of particles from A-blocks  to B-blocks (see main text). In Section~\ref{ssec:bcmodel}, we introduce an effective model where clusters are artificially bound by a negative potential well. In addition, we compare the original model of the main text to the effective model with the mass of all clusters set equal. Numerical DMRG simulations show no significant discrepancies between the original and the effective models for sufficiently strong interactions, justifying the application of both approximations ({\it i}) and ({\it ii}) in our low-energy field theory. 

\subsection{Cluster Stability from the Classical Energy}
\label{ssec:classical}

\paragraph{Tunneling energy barriers.} - 
As an example, we consider the case $r_c=4$ and $\bar n = 1/4$ of Fig.1(c) of the main text. There, an A-block is formed by two particles followed by four empty sites, while a
B-block is formed by a single particle followed by four empty sites (see main text). All configurations which can be written as a sequence of A-blocks and B-blocks with the constraint $N(\mathrm{B}) = 2 N(\mathrm{A})$, are ground-states with identical energies [here, $N(\mathrm{A})$ and $N(\mathrm{B}$) are the number of A- and B-blocks, respectively]. 

\begin{figure}[bt]
\centerline{\includegraphics[width=0.95\columnwidth]{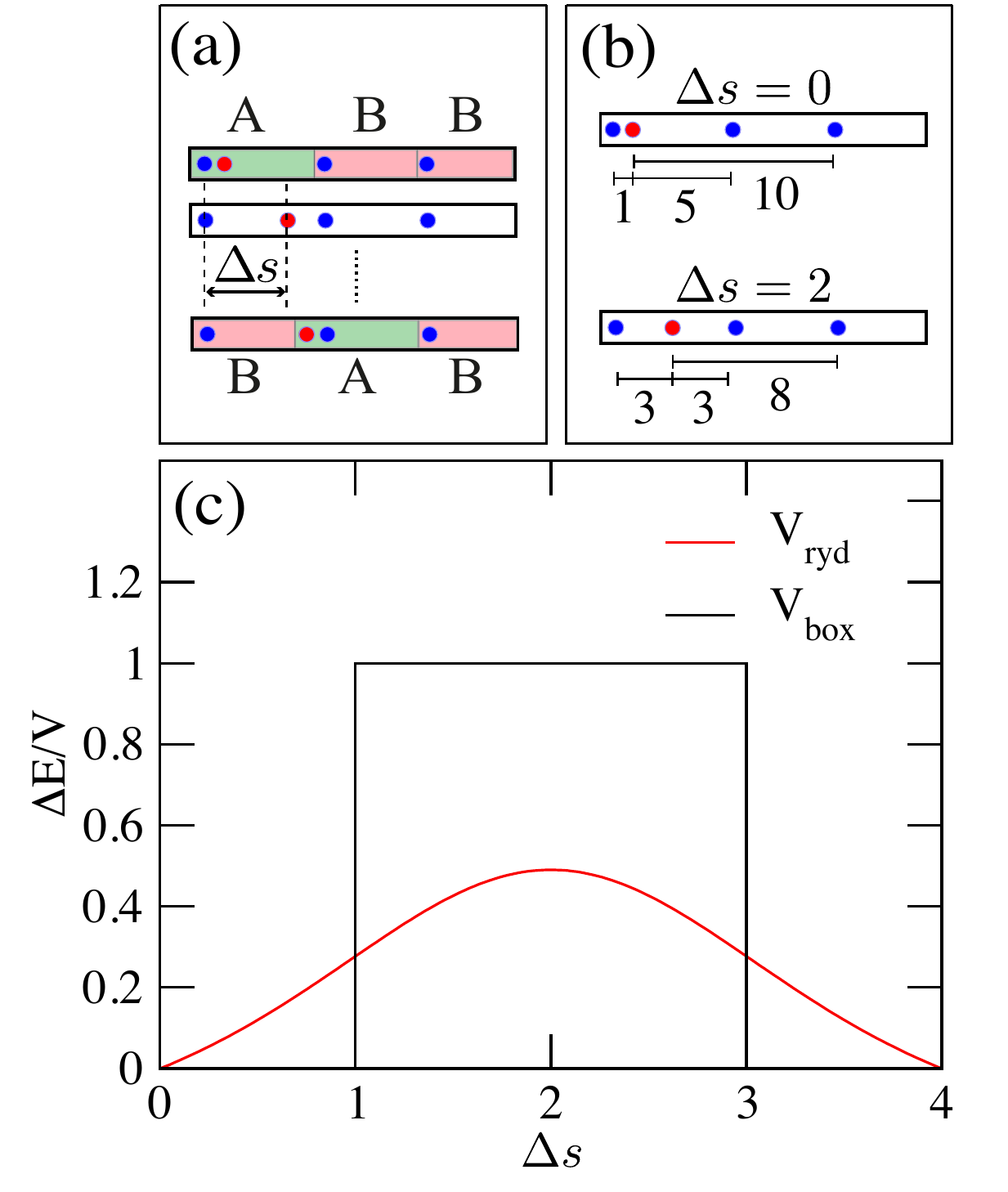}}
\caption{In the classical system, the ground-state is degenerate and separated by a large energy barrier from excited states.
(a) The physical move associated with an A- and B-block exchange is the tunneling of one cluster-particle (red sphere) from a doubly-occupied A-block (green) to a neighboring singly-occupied B-block (red). $\Delta s$ is defined as the displacement of a cluster-particle with respect to its ground-state original position (defined as $\Delta s = 0$). $\Delta s=4$ is again a ground-state configuration and corresponds to the A- and B-block exchange process completed. (b) For $r_c=4$, the ground-state and the maximally excited $\Delta s = 2$ configurations are shown.  Other excited states are $\Delta s = 1,3$ (not shown). (c)
At $\Delta s = 2$, the tunneling energy barrier of the simplified box potential (black) is roughly a factor of 2 larger with respect to the Rydberg-dressed potential barrier (red).}
\label{fig:mr4}
\end{figure}

As described above, each block contains at least one particle [blue sphere in the sketch of Fig.~\ref{fig:mr4}(a)]. For later convenience, we now define a {\it cluster-particle} as the additional ("second") particle which belongs to a cluster of the A-block [red sphere in Fig.~\ref{fig:mr4}(a)].  In the lowest-energy configuration the two particles in the A-block sit next to each other. As sketched in Fig.~\ref{fig:mr4}(a)-(b),  this corresponds to a displacement $\Delta s=0$ (in units of lattice constants $a$) of the cluster-particle from its ground-state position. Because of the finite-range interactions of our models, any finite displacement $\Delta s \neq 0$ of the cluster-particle costs energy. In particular, for the case of Fig.~\ref{fig:mr4} a displacement of $\Delta s = 4$ results in an effective {\it exchange} of the A-block with a neighboring B-block [see panel (a)]. We define the energy cost associated with this exchange process as {\it tunneling energy barrier} (where the term "tunneling" is inspired by the corresponding quantum mechanical exchange process).

\begin{figure}[t]
\centerline{\includegraphics[width=0.95\columnwidth]{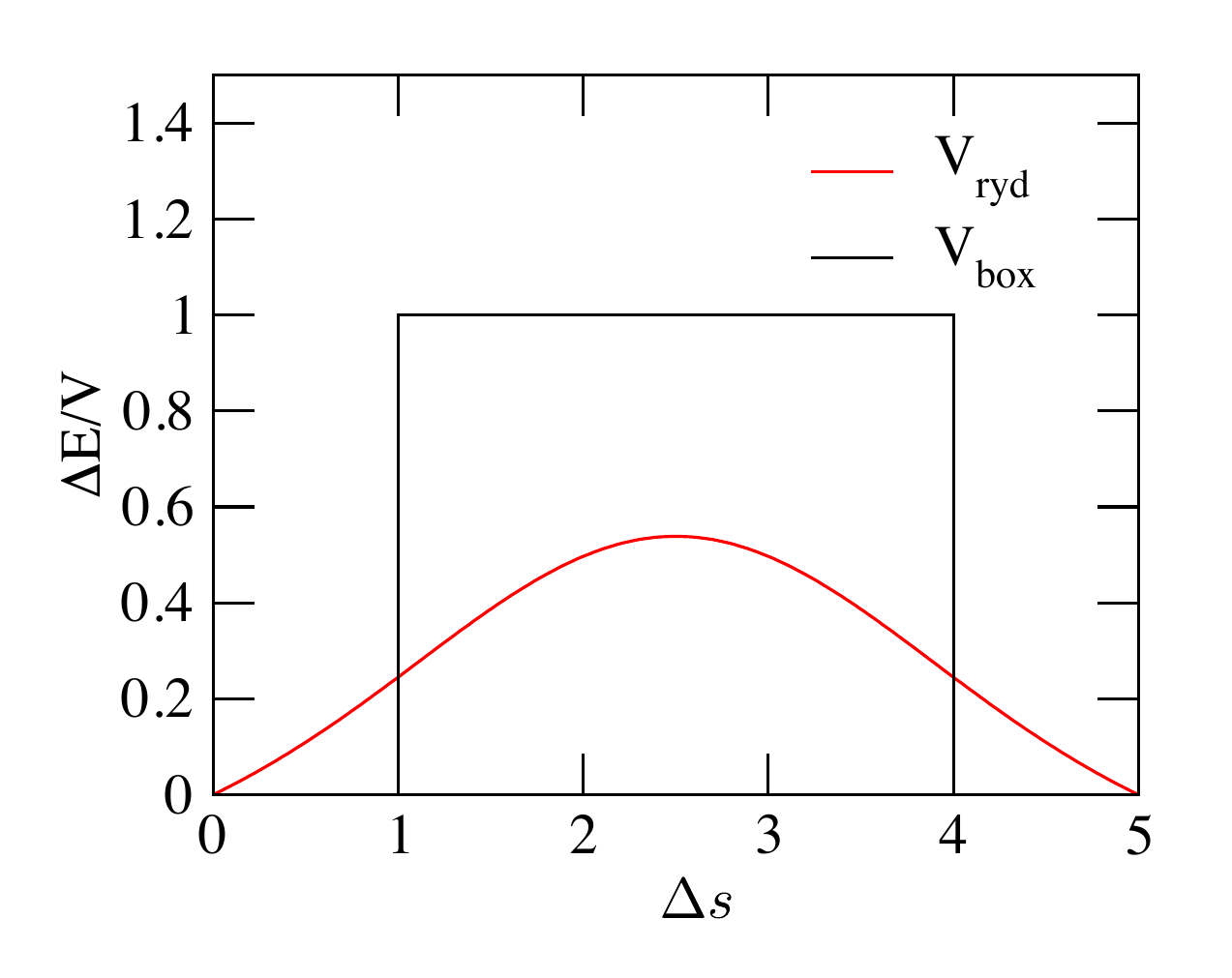}}
\caption{Tunneling energy barriers of the box and Rydberg-dressed potentials for $r_c=5$, as a function of several distances $\Delta s$ of the cluster-particle from its original ground-state position. $\Delta s = 5$ is again a ground-state configuration. For $\Delta s =2$ and $3$, the tunneling energy barrier of the box potential (black) is roughly a factor of 2 larger with respect to the Rydberg-dressed potential barrier (red).}
\label{fig:mr5}
\end{figure}

Fig.~\ref{fig:mr4}(c) compares the tunneling energy barrier of the simplified box potential [Eq. (1) of the main text with $t=0$] with the corresponding one for the Rydberg-dressed potential $V_{\mathrm{ryd}}(i,j) =  V_{\mathrm{ryd}}/\{1+[(i-j)a/r_c]^6\}$. We note that in the latter case, the barrier is considerably smaller than in the case of the box potential, which immediately suggests that stronger interactions will be needed in the case of Rydberg-dressed interactions to obtain the same clustering effects of the case of a box potential. We quantify this feature by computing explicitly the tunneling energy barrier for $(i)$ the  box potential and $(ii)$ the full Rydberg-dressed potential. $(i)$ In the case of the box potential, the only energy contribution to the ground-state configuration is the pair interaction of the two particles in the A-block, $E_{\Delta s = 0} = V$. The energy of the system for $\Delta s=2$ is instead $E_{\Delta s = 2} = 2 V$, as the cluster-particle interacts with two neighbors [Fig.~\ref{fig:mr4}(b)]. The resulting tunneling energy barrier is thus $\Delta E = E_{\Delta s = 2}- E_{\Delta s = 0}= V$ [black line in Fig.~\ref{fig:mr4}(c)]. $(ii)$ In the case of the Rydberg-dressed interaction, the energy of a certain configuration is in general the sum of the interaction with {\it all} other particles, that is, $E_{\Delta s} = \sum_{i \neq j} V_{\textrm{ryd}}(i,j)=\sum_{i \neq j} V_{\textrm{ryd}}(i-j)$. For the case of Fig.~\ref{fig:mr4}(b), the tunneling energy barrier is then $\Delta E = E_{\Delta s=2} - E_{\Delta s = 0} = V_{\textrm{ryd}}(3)+V_{\textrm{ryd}}(3) + V_{\textrm{ryd}}(8) - [V_{\textrm{ryd}}(1) +V_{\textrm{ryd}}(5)+V_{\textrm{ryd}}(10)] = (1.713 - 1.211) V_{\textrm{ryd}} = 0.502 \, V_{\textrm{ryd}}$ [maximum of the red curve in Fig.~\ref{fig:mr4}(c)]. This latter value is about a  factor of 2 smaller than for the case of the box potential. As noted above, in the quantum mechanical problem ($ t\neq 0$) this difference is at the origin of the requirement $V_{\mathrm{ryd}}/V \gtrsim 2$ for the agreement in observables like $g_2(x)$ and $S(k)$ between the two cases (see Fig.~4 of the main text).

Similarly to the discussion above, Fig.~\ref{fig:mr5} shows the tunneling energy barriers for the case with $r_c=5$. Here, an A-block is formed by two particles followed by five empty sites, whereas a B-block is formed by a single particle followed by five empty sites. In order to determine an exchange between an A-block and a neighboring B-block, a cluster-particle has thus to move a distance $\Delta s = 5$. Fig.~\ref{fig:mr5} shows that the tunneling energy barrier is here again reduced roughly by a factor of 2 in the case of the  Rydberg-dressed potential with respect to the case of the box potential.

We conclude this subsection by remarking that the existence of the tunneling energy barriers discussed above is at the origin of the strong suppression of fluctuations of the cluster-particle position for finite (but large) ratios of $V/t$ in the quantum regime ($t\neq 0$). This fact justifies our assumption ($i$) in the derivation of the low-energy field theory.

\begin{figure}[t]
\centerline{\includegraphics[width=0.95\columnwidth]{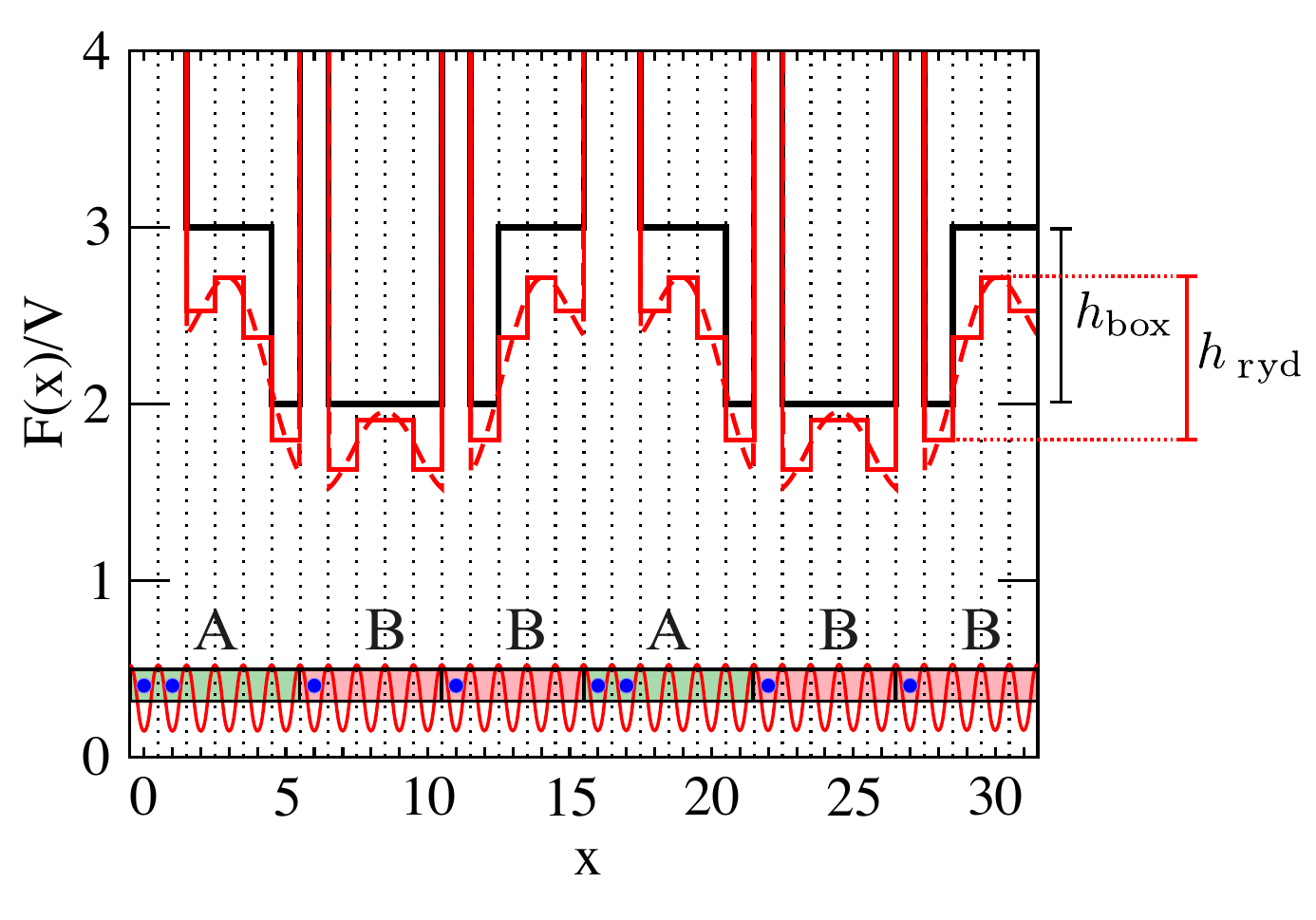}}
\caption{Comparison of the ground-state energy landscape $F(x)$ (see text) at filling $\overline{n}=1/4$ and $r_c=4$ for the box potential (black), the discrete (red) and free-space (red, dashed) Rydberg-dressed potentials. Noticeably, all landscapes have similar barrier heights, i.e. $h_{\mathrm{ryd}}=0.919 \, h_{\mathrm{box}}$. However, the "multi-step-like" structure of the Rydberg-dressed cases render cluster structures less stable with respect to the box case. We note that the physical interpretation of these barrier heights is different with respect to the tunneling energy barrier of Fig.~\ref{fig:mr4}: here, $F(x)/V$ denotes the energy cost of introducing a single test particle on top of a classical ground-state configuration, while $\Delta E/V$ in Fig.~\ref{fig:mr4} describes the energy cost of moving a particle within the cluster ground-state manifold (no additional particle is added).  }
\label{fig:r4}
\end{figure}

\begin{figure}[h]
\centerline{\includegraphics[width=0.95\columnwidth]{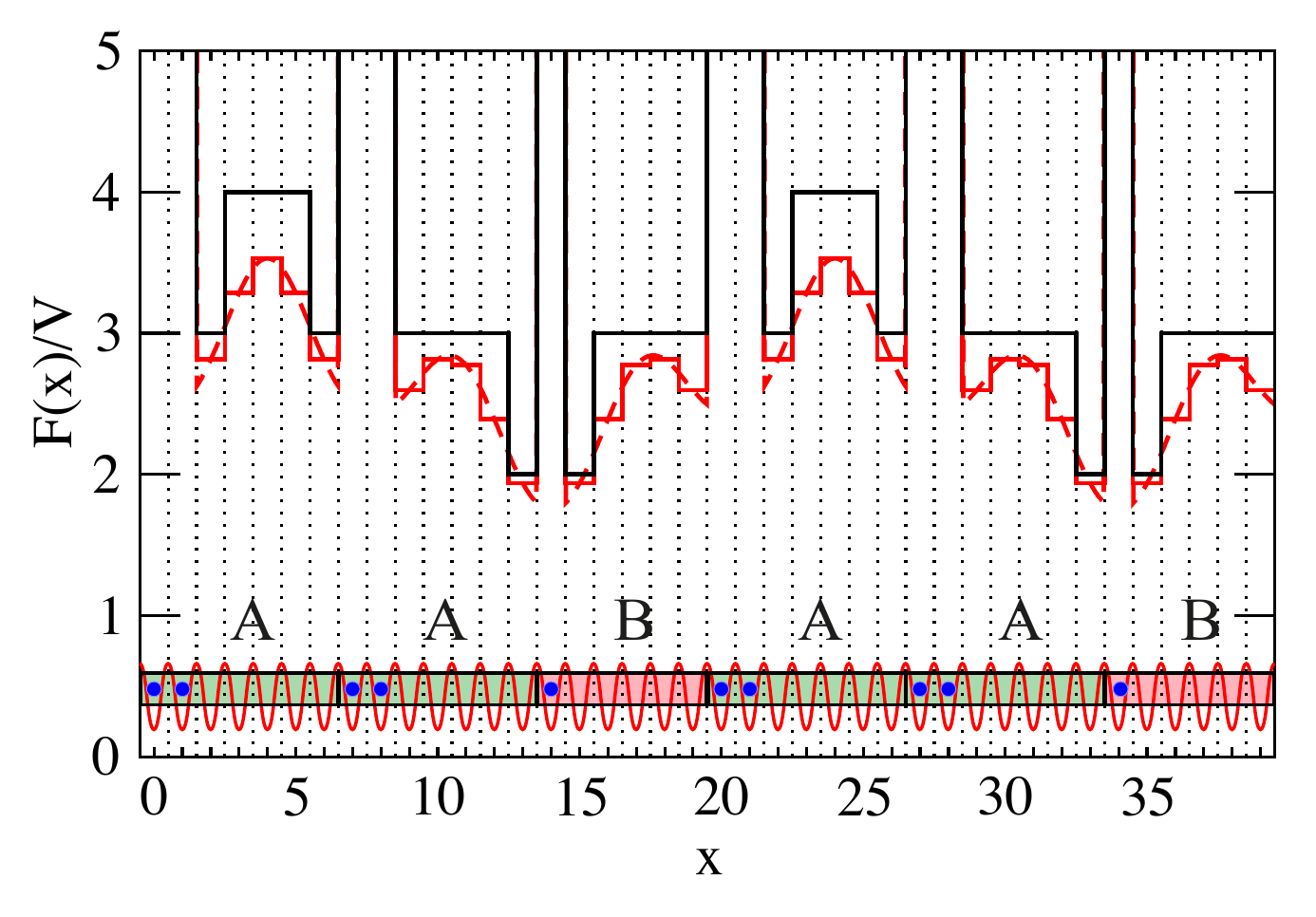}}
\caption{Comparison of the ground-state energy landscape at filling $\overline{n}=1/4$ and $r_c=5$ for the simplified box potential (black), the discrete (red) and free-space (red, dashed) Rydberg-dressed potentials. In analogy to the $r_c=4$ case (Fig.~\ref{fig:r4}), barrier heights are very similar for landscapes of different potentials.}
\label{fig:r5}
\end{figure}

\paragraph{Energy landscapes of test particles.}-
We conclude the discussion of classical energies by considering the {\it energy landscape} $F(x)$ for the two models above. $F(x)$ is defined as the energy experienced by a test particle added at position $x$, while keeping all other particles in the original ground-state configuration. Different with respect to the case of the tunneling energy barrier discussed above, we thus now investigate a {\it doped} system. The interpretation of the energy landscape is as follows: if a test particle is added to the system, it will be attracted by the local minima of $F(x)$. For example, in Fig.~\ref{fig:r4}, the test particle in the Rydberg-dressed case will sit exactly in correspondence of singly-occupied B-blocks, precisely at positions $x=7,10,23,26$. For the box case, the flat shape of the interaction potential increases the degeneracy of the number of local minima, which are now located at positions $x=5,7,8,9,10,12,21,23,24,25,26,28$. Similar results for $r_c=5$ are illustrated in Fig.~\ref{fig:r5}.

The effective attractive potential shown in Fig.~\ref{fig:r4} and \ref{fig:r5} for test particles close to the B-blocks is typical of clustering in classical systems. The analysis presented here confirms this effect for our model Hamiltonians in the classical regime, and constitutes a basic first step to guide the investigation of the quantum regime provided in the main text. We note however that the effective potential felt by the test particle (depicted in Fig.~\ref{fig:r4} and \ref{fig:r5}) does not correspond to the tunneling energy barrier related to the block exchange discussed in the previous section. Therefore, it does not give direct information on the stability of the clustered states once the potential is changed from the box-shaped case to the Rydberg-dressed case.

\subsection{Bound Clusters Model}
\label{ssec:bcmodel}

\begin{figure}[h!]
\centerline{\includegraphics[width=0.9\columnwidth]{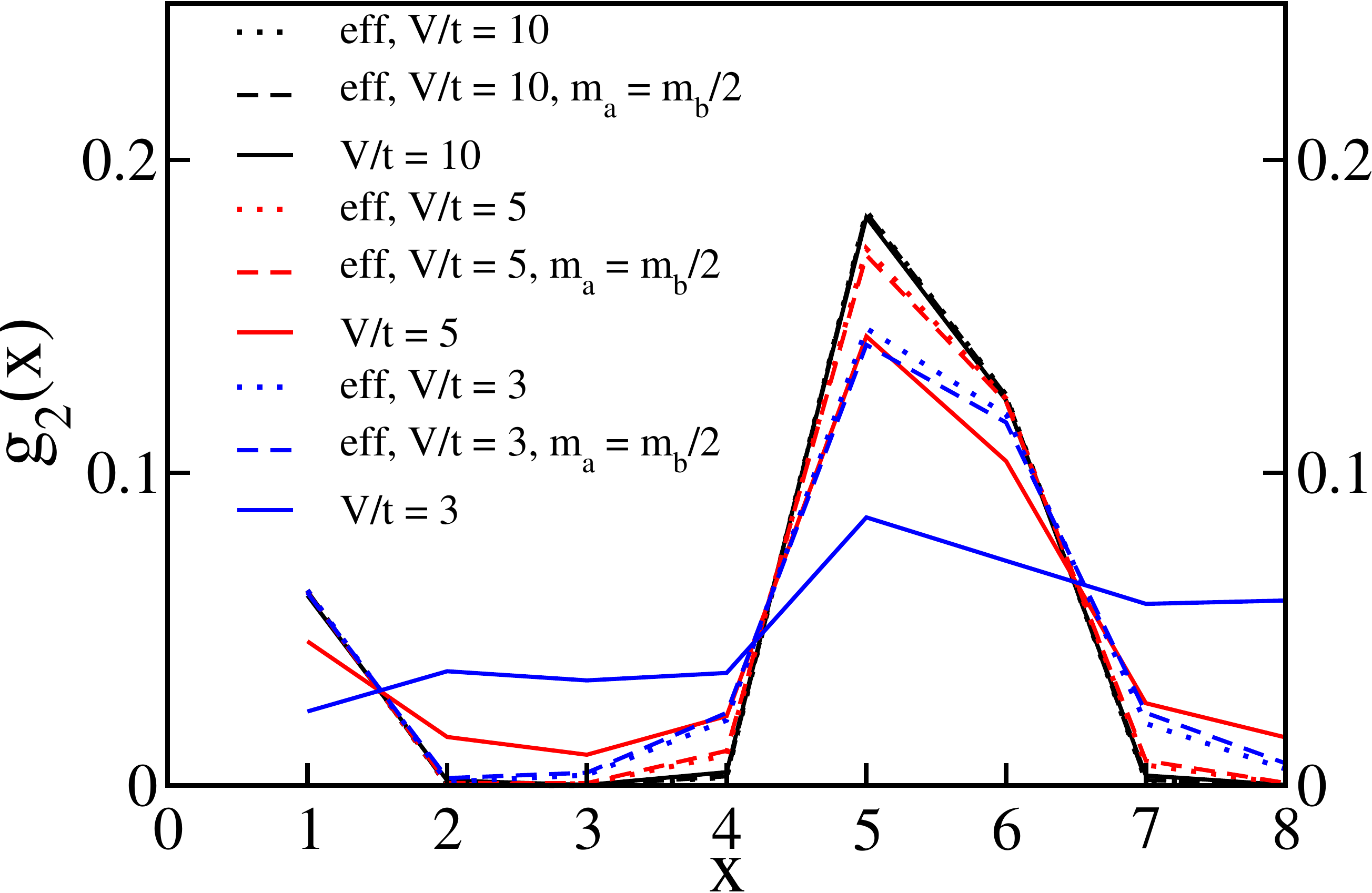}}
\caption{Comparison of the density-density correlation functions $g_2(x)$, for the effective model of Eq.~\ref{ham_eff} (dotted lines), the effective model with modified masses (dashed lines) and box potential model of Eq.(1), main text (solid lines).
Remarkably, for $V/t=10$ (black lines), the three correlators agree with very high accuracy.}
\label{fig:eff}
\end{figure}

In the following, we study the range of validity of assumption ($i$) of our low-energy theory with an effective model where clusters are 'artificially' built up.
If we consider the illustrative example of $r_c = 4$ for $L = 16$, the classical ground-state is characterized by the configuration ABB (and associated block permutations). \\
In order to properly simulate tightly-bound clusters, we introduce an effective Hamiltonian which reads:
\begin{eqnarray} \label{ham_eff}
H_{\mathrm{eff}} = &U& \sum_i n_{i, a}n_{i+1, a} + V\sum_{i} \sum_{\ell=2}^{r_c} n_{i, a} n_{i+\ell, a} + \nonumber \\
&+& V\sum_{i} \sum_{\ell=1}^{r_c} n_{i, b} n_{i+\ell, b} + V\sum_{i,\sigma \neq \sigma^{\prime}}\sum_{\ell=1}^{r_c} n_{i,\sigma}n_{i+\ell,\sigma^{\prime}} +  \nonumber \\
&-&\sum_{i,\sigma} t_{\sigma} \, (b^{\dagger}_{i,\sigma} b_{i+1,\sigma}+\textrm{h.c.}).
\end{eqnarray}
Here, both $\sigma$ and $\sigma^{\prime}$ label two different species of particles $a$ and $b$, in analogy to the building block labels A and B to which they belong, respectively.
Accordingly to the main text formalism, $b^{\dagger}_{i,\sigma}$ ($b_{i,\sigma}$) is the creation (annihilation) operator of the $\sigma$-species hard-core boson - spinless fermion - at the $i$-th site,
whereas $n_{i,\sigma}=b^{\dagger}_{i,\sigma}b_{i,\sigma}$ is the $\sigma$-species corresponding number operator. $t_{\sigma}$ represents the amplitude
of the $\sigma$-species hopping term. Unless otherwise stated, we set $t_{a} = t_{b} = t$.\\
The term with $U<0$ ($|U|\gg t$) represents
a strong \textit{attractive} nearest-neighbor interaction between $a$-species particles, which clearly favors cluster formation.
As expected, the Hamiltonian of Eq.~\ref{ham_eff} simulates the ground-state of our model for large $V/t$.\\
In Fig.~\ref{fig:eff}, we compare numerically computed density-density correlation functions of the simplified box potential model [Eq.(1), main text; $g_2(x) = \langle n_{1} n_{1+x} \rangle - \overline{n}^2$] with the one of the effective model [Eq.~\ref{ham_eff}, $g^{\mathrm{eff}}_2(x) = \sum_{\sigma, \sigma^{\prime}} \langle n_{1,\sigma} n_{1+x, \sigma^{\prime}} \rangle - \overline{n}^2$],
for $U/t = -20$ and several values of $V/t$.
For sufficiently strong interactions, $g_2(x)$ and $g^{\mathrm{eff}}_2(x)$ are in excellent agreement (see, e.g., $V/t=10$, black lines).
This result justifies the approximation, in our low-energy field theory,
of neglecting relative correlations between $a$-species particles underlying each cluster, which in turn are assumed to be single objects with no internal degrees of freedom. \\
At lower interactions (e.g. $V/t=3$, blue lines), the agreement reduces substantially,
suggesting that the effective model overestimates "clustering" respect to the original one, which is non-negligibly affected, in this regime of parameters,  by internal (cluster) relative modes.

\subsubsection{Unequal Hopping Amplitude Approximation}

Within this approximation, we halve the mass of the $a$-species particles respect to the $b$-species ones,
ensuring that the resulting two-$a$-particle cluster has the same effective hopping amplitude of the $b$-species single particles
(as $t_{\sigma} \propto 1/m_{\sigma}$, with $m_{\sigma}$ the mass of the $\sigma$-species particle, we obtain $t_a = 2t_b$). This constraint is introduced to study the validity of the assumption $(ii)$ in our low-energy field theory of neglecting the mass difference between A- and B-type clusters. \\
Again, DMRG results support the aforementioned approximation for $V/t_b=10$, the overlap of $g_2(x)$ and $g^{\mathrm{eff}}_2(x)$ being even more precise than in the case $t_a = t_b = t$, as expected.  \\

\section{Numerical Results}

\subsection{Technical Details}

In this section, we provide additional details on the DMRG calculations
presented in the main text. \\
In all simulations we considered a filling $\overline{n} = 3/4$, unless otherwise stated. \\
In order to avoid boundary effects, which can be significant due to the finite range of interactions, we perform finite-size scaling in systems as large
as $L=64$ sites with periodic boundary conditions, keeping up to $1800$ optimized states per system-block and performing $5$ sweeps. This ensures truncation errors in the DMRG procedure of the order $10^{-6}$ or less.

A first remark concerns the systems sizes
considered: while in the regime $r_c\leq r^{\star}$ all lengths $L=4\ell, \ell\in\mathbb{N}$
are appropriate for finite-size scaling purposes, this is not true anymore
in the region where effects due to the cluster structure could emerge.
In the latter cases, one has to consider systems sizes which can arrange
a proper cluster structure in the corresponding classical limit.
For $r_c=4$, this imposes $L=16, 32, 48, 64$, while for $r_c=5$,
$L=20, 40, 60$ (the upper bound is of course limited by computational time). This choice ensures an appropriate
scaling procedure, although limited to a small number of available
system sizes.

\begin{figure}[h!]
\centerline{\includegraphics[width=0.85\columnwidth]{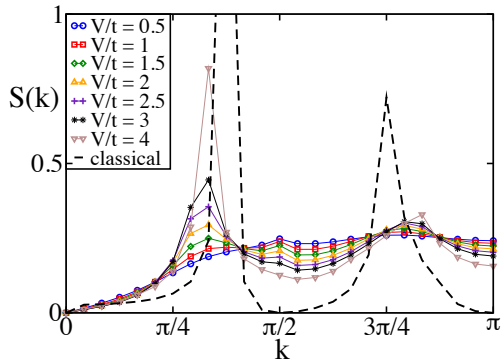}}
\caption{Evolution of the static structure factor as a function of
$V/t$ from the Luttinger liquid to the cluster Luttinger liquid phase at $r_c=4$. The dashed
black line represents the classical prediction.}
\label{fig:Sk}
\end{figure}

Classical simulations are performed within the
same system sizes, although in this case calculations could be extended on
much larger $L$. A typical comparison for the static structure factor $S(k)$,
illustrating the emergence of anomalous peaks in analogy to
the classical prediction, is presented in Fig.~\ref{fig:Sk}  for several values of
$V/t$ at fixed $r_c=4$. The small deviation $\delta k = 2\pi/L$ between the DMRG-evaluated position of the critical momentum $k_c$ of the peak in $S(k)$ and the
corresponding classical value will be subject of future investigations.

\subsection{Estimates of the Luttinger parameter}

Here we present three different methods we used for the evaluation of the Luttinger parameter
$K$ (for a complete review, see Ref.~\onlinecite{GIAMARCHI2003}). In particular, we show that for $r_c>r^{\star}$ and $V/t \gtrsim 1$, the estimates are inconsistent, signaling the breakdown of the standard TL paradigm. 

\subsubsection{Static structure factor method}

The Luttinger liquid theory predicts that the finite-size static structure factor $S_L(k)$ scales linearly at small momenta $k$ according to:
\begin{equation}\label{static}
S_L(k) \simeq k \dfrac{2K_L}{\pi}
\end{equation}
where $K_L$ is the size-dependent Luttinger parameter.
In a periodic chain with $L$ sites, the smallest non-vanishing momentum of the discrete set $k = \tfrac{2\pi p}{L}$ ($p \in [0; L-1]$) is $2\pi/L$; rewriting Eq.~\ref{static}
as

\begin{equation}
K_L = \dfrac{LS_L(2\pi /L)}{4},
\end{equation}
it is possible to compute $K_L$ by inserting the Fourier transform of the DMRG-evaluated density-density correlation functions $g_2(x)$
(see main text). Finally, using standard finite-size scaling procedures, we extrapolate the value of the Luttinger parameter in the thermodynamic limit,
$K \equiv \lim_{L\rightarrow\infty} K_L$.

\subsubsection{Level spectroscopy method}
For $c=1$ conformal field theories, the ground-state energy density $\epsilon_{gs}(L)$ scales as

\begin{equation}\label{gs}
\epsilon_{gs}(L) = \epsilon_0 + \dfrac{\pi v}{6L^2} + O(1/L^3),
\end{equation}
where $\epsilon_0 = \lim_{L\rightarrow\infty} \epsilon_{gs}(L)$. From Eq.~\ref{gs}, it is possible to extract the sound velocity $v$.
Given that
\begin{equation}
K_L = \dfrac{\pi v}{2 G(L) L} + O(1/L^2),
\end{equation}
where
\begin{equation}
G(L) = \dfrac{E^{N+1}_{gs}(L) + E^{N-1}_{gs}(L) - 2E^{N}_{gs}(L)}{2},
\end{equation}
is the charge gap ($E^N_{gs}(L)$ representing here the ground-state energy of $N$ particles in $L$ sites), and performing finite-size scaling as above, we obtain a second \textit{independent}
estimate of the Luttinger parameter $K$.

\subsubsection{One-body density matrix method}
According to Ref.~\onlinecite{CAZALILLA2004}, in a system with periodic boundary conditions described by a standard Luttinger liquid theory,
the one-body density matrix $B_{\ell j}$ is parametrized by $K$ as follows:

\begin{eqnarray}\label{one}
B(\ell- j) &= & \overline{n} \cdot \left(\ \dfrac{1}{\overline{n} \, d_{\ell j}(L)}\right)^{\frac{1}{2K}}\times\\
&\times& \left[ c_0 + \sum_{m=1}^{\infty} c_m \cdot \left(\dfrac{1}{\overline{n} \, d_{ \ell j}(L)}\right)^{2m^2K} \cos(2 \pi \overline{n} m x) \right] \nonumber
\end{eqnarray}
where $c_m$ are model-dependent coefficients and $d_{\ell j}(L)= (L/\pi) \lvert \sin[\pi (\ell -j)/L] \rvert$ is the so called cord-length.
We performed a fit of the DMRG computed one-body density matrixes with Eq.~\ref{one}, retaining two harmonics in the series ($m=1,2$).
We verified that the error associated to this approximation is smaller than the desired precision in $K$ ($\sim 10^{-3}$).
The four free parameters of the fit are thus $c_0, c_1, c_2$ and $K$, the latter being the one of our interest. This is thus the third independent estimate of $K$,
which can be compared with the structure factor and level
spectroscopy one, in order to confirm the departure from the Luttinger liquid paradigm in a certain regime of parameters of the phase diagram (see Fig. 3, main text).

\subsubsection{Comparison of the different Luttinger parameters}
\begin{figure}[h!]
\begin{center}
\includegraphics[width=0.46\columnwidth]{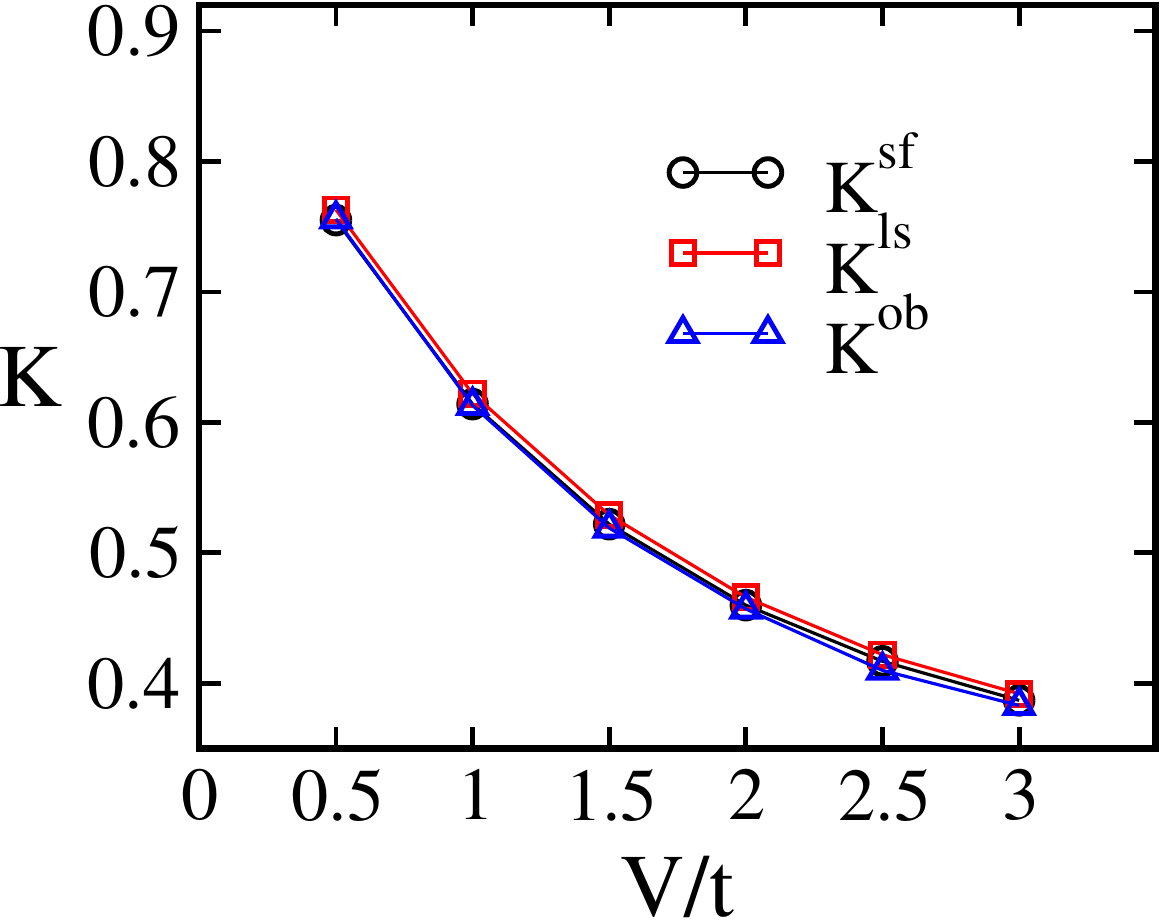}
\includegraphics[width=0.46\columnwidth]{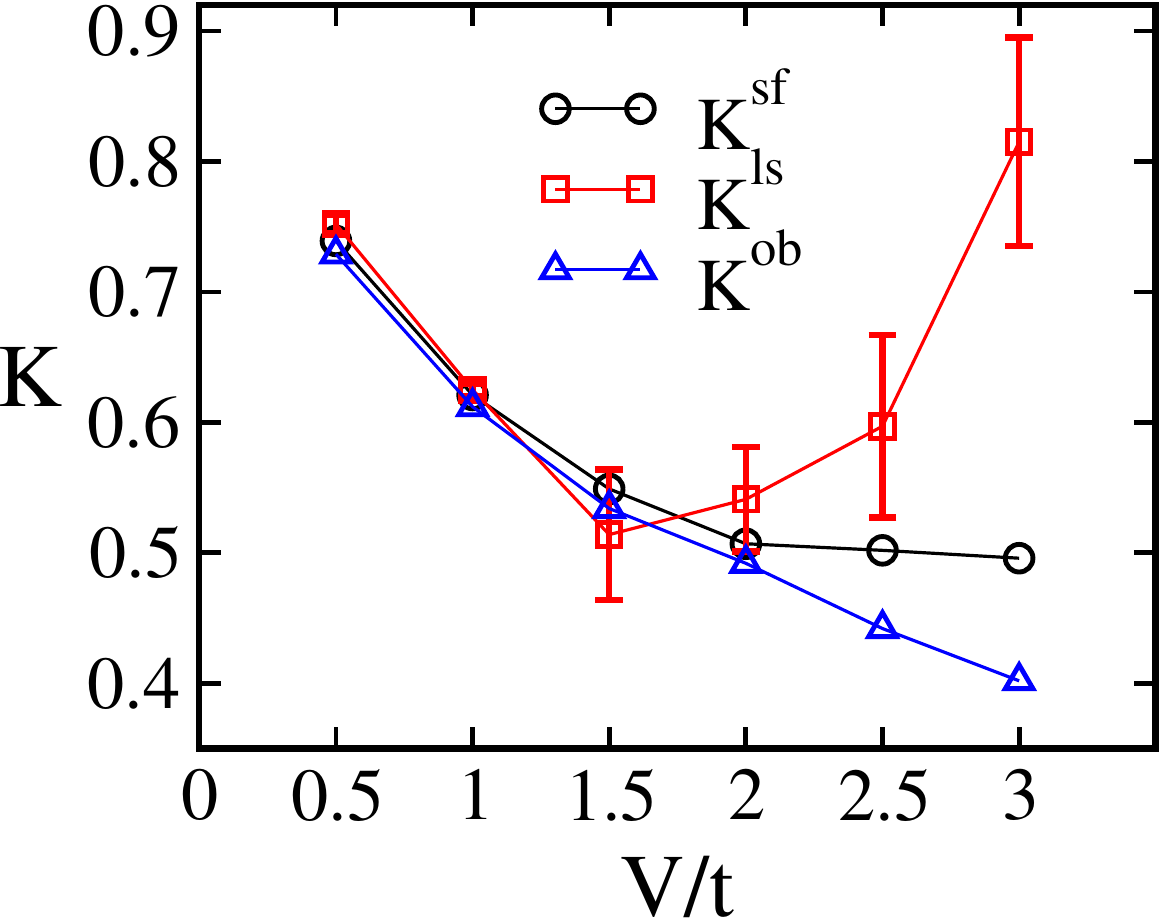}
\caption{Estimate of $K$ by means of independent methods.
Left panel: $r_c=2$. All estimates agree, signaling the stability of the Luttinger liquid phase.
Right panel: $r_c=4$. Except at small interactions, the estimates are not consistent,
signaling the breakdown of the standard Luttinger liquid scenario.}
\label{fig:Kpar}
\end{center}
\end{figure}

In Fig.~\ref{fig:Kpar}, we show a comparison of the Luttinger parameters $K$ calculated using the structure factor, level spectroscopy and one-body density matrix methods
for two illustrative examples: the (usual) Luttinger liquid phase ($r_c=2$, left panel) and the cluster Luttinger liquid phase ($r_c=4$, right panel).

As evident, for $r_c=2$, all three methods agree very well, the related accuracy on the absolute value of $K$ being higher than $10^{-2}$.
Contrarily, for $r_c=4$ and sufficiently large interactions, a discrepancy between the three methods manifests, which can not be justified exclusively by the
enlarged uncertainty in the fit procedure (error bars, if not shown, are smaller than the symbol size).

\subsection{One-body density matrix}
\begin{figure}[h!]
\centerline{\includegraphics[width=0.75\columnwidth]{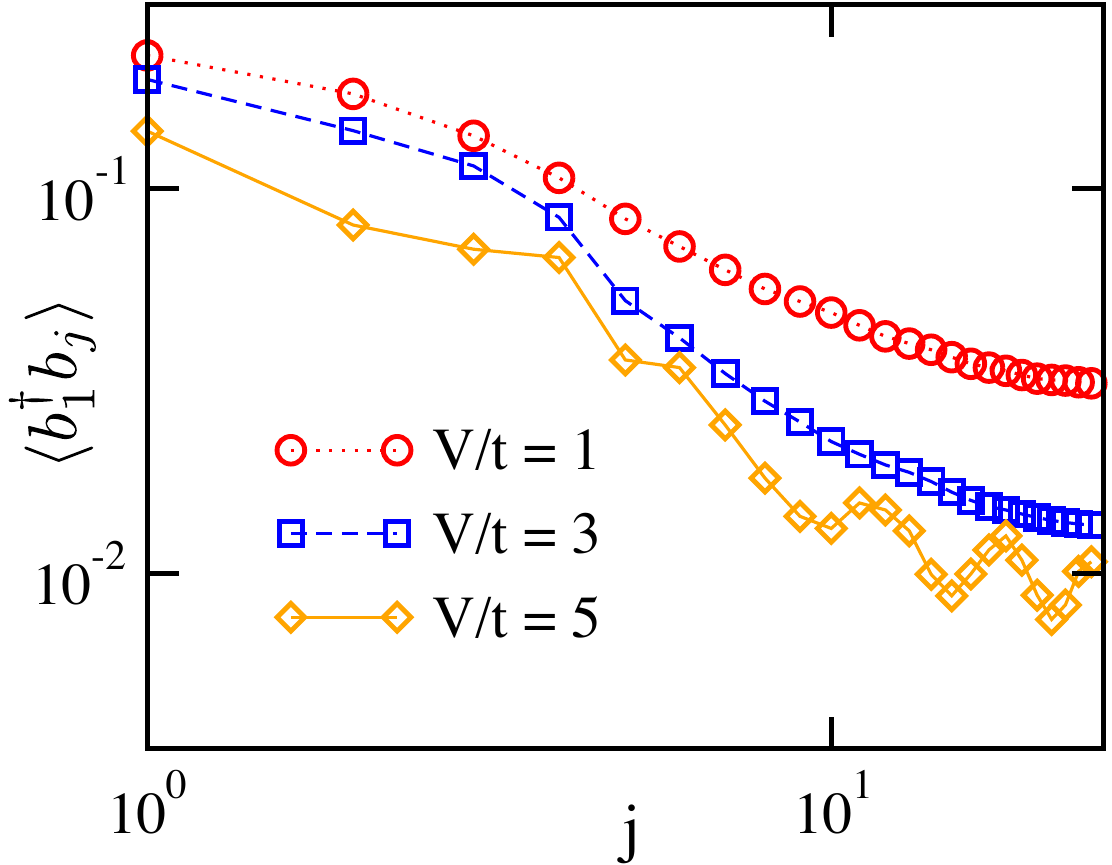}}
\caption{Log-log plot of the one-body density matrix $B(1-j)$ for $\overline{n}=3/4$, $r_c =4$ and $L=48$ sites, at different interactions. For $V/t=5$ (orange diamonds), the large asymptotic oscillations determine the emergence of the anomalous peak in
$n(k)$ at non-vanishing momentum (see Fig. 2(c), main text).}
\label{fig:ob}
\end{figure}

The anomalous peak in the momentum distribution $n(k) = \sum_{\ell ,j} e^{ik( \ell -j)} B(\ell - j)/L$ visible in Fig. 2(c) of the main text at non-zero $k$-vector for $r_c = 4 > r^{\star}$,
is associated to large oscillations in the corresponding one-body density matrix $B(\ell - j)$. In Fig.~\ref{fig:ob},
we show the emergence of these asymptotic behavior for sufficiently large interactions ($V/t=5$, orange diamonds).

\subsection{Excitation spectrum}
\begin{figure}[h!]
\centerline{\includegraphics[width=0.75\columnwidth]{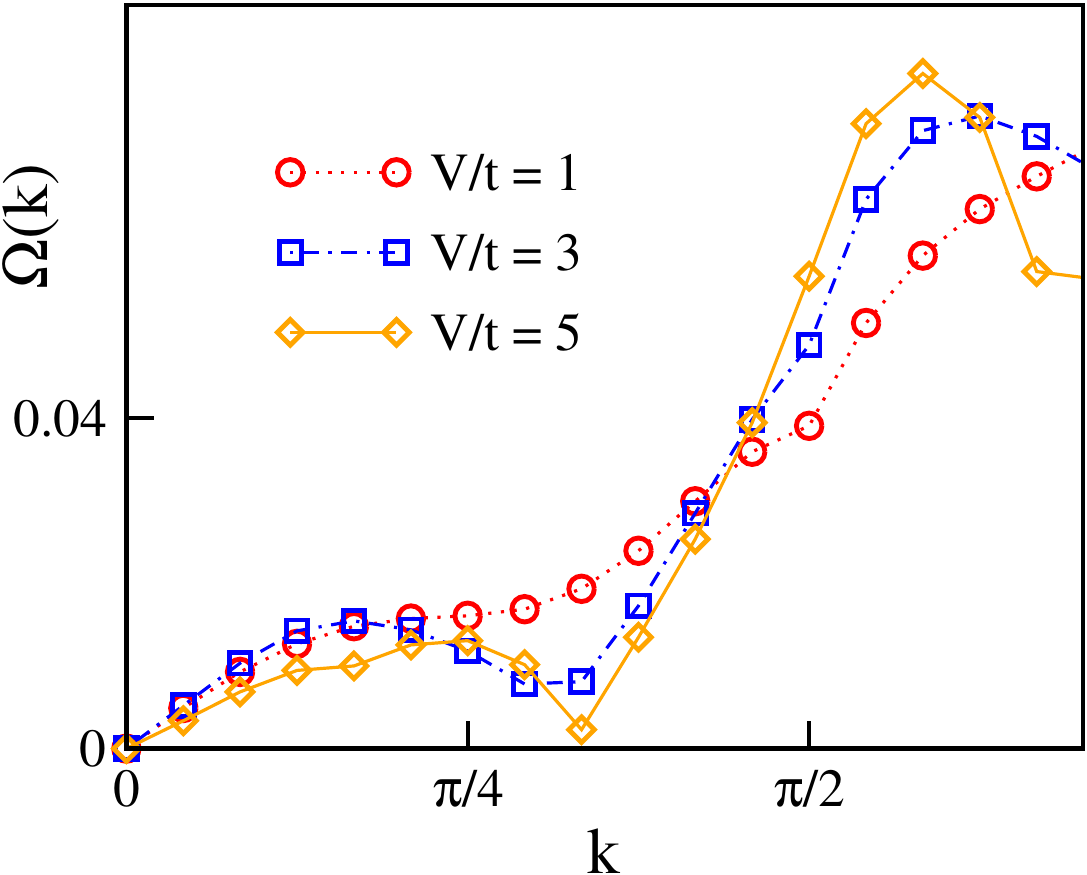}}
\caption{Excitation spectrum for $\overline{n}=3/4$, $r_c =4$ and $L=48$. Increasing interactions $V/t$, a cluster-like roton minimum instability appears with a null roton gap.}
\label{fig:ex}
\end{figure}

The excitation spectrum $\Omega(k)$ in the cluster Luttinger regime features a cluster-type roton minimum at $k_c<\pi/2$. In Fig.~\ref{fig:ex}, we show the formation and evolution of the roton
instability for $r_c = 4$ and increasing interactions. The roton critical momentum $k_c$ is identical to both the static structure factor and momentum distribution peaks (Figs. 2(a) and 2(c) of the main text, respectively).

\subsection{BKT transition at $r_c=3$}

\begin{figure}[h!]
\centerline{\includegraphics[width=0.8\columnwidth]{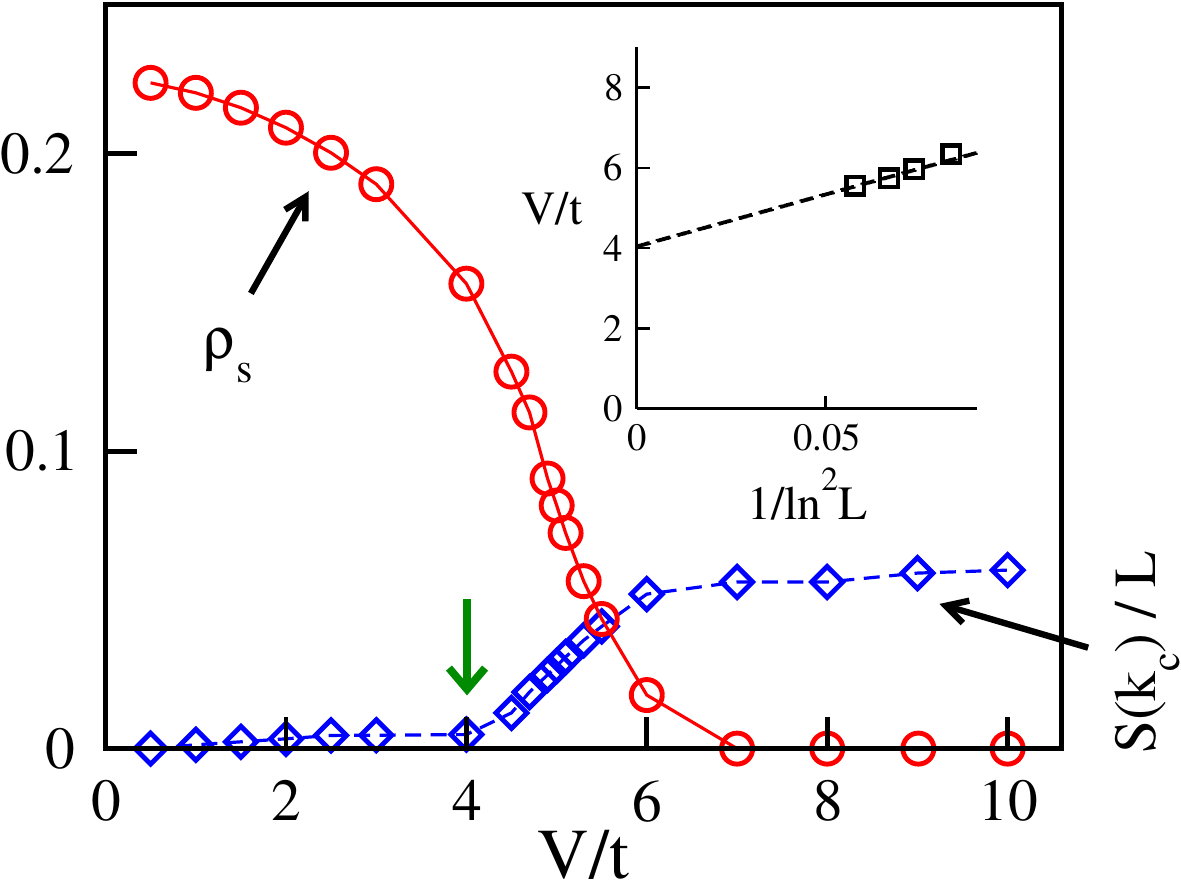}}
\caption{Numerical results on the BKT transition at $r_c=3$. Plot of the superfluid stiffness $\rho_s(48)$ (red circles) and $S_{48}(\pi/2)/48$ (blue diamonds), increasing interactions from $V/t=0.5$ to $10$. Inset: the BKT predicted linear scaling (see text) is recovered via a fit (dotted black line) of
the numerical results for the quantity $(V/t)_c(L)$ at different system sizes.}
\label{fig:BKT}
\end{figure}

In order to estimate quantitatively the critical interaction $(V/t)_c$ of the expected phase transition from a Luttinger liquid to a (classical) crystal
at $r_c = r^{\star} = 3$ (see Fig.3(a) of the main text), we considered the following criterion: we calculated, for
different system sizes, the interaction $(V/t)(L)$ at which the superfluid stiffness in its discretized form (see Ref.~\onlinecite{DENG2009})

\begin{equation}\label{sup_stiff}
\rho_s (L) = \dfrac{L}{t}\dfrac{\partial^2 E^{\theta}_{gs}(L)}{\partial \theta^2} \simeq \dfrac{L}{t} \dfrac{E^{\theta}_{gs}(L) - E^{0}_{gs}(L)}{\Delta\theta^2},
\end{equation}
is smaller than a certain threshold (we choose $\rho_s (L) < 0.02$).
Here $E^{\theta}_{gs}(L)$ is the ground-state energy of a system with boundary conditions twisted by an angle $\theta$ respect to periodic
boundary conditions (defined as $\theta \equiv 0$).
Following the procedure illustrated in Ref.~\onlinecite{DENG2009}, we checked numerically that $\rho_s(L)$, close to $(V/t)_c(L)$, does not vary significantly
when small (ideally infinitesimal) $\delta\theta$ variations are replaced, as in Eq.~\ref{sup_stiff}, with larger ones $\Delta\theta = \pi$
($\theta = \pi$ representing anti-periodic boundary conditions).
According to the Berezinskii-Kosterlitz-Thouless universality class of phase transitions, the correlation length $\xi$ of the system scales as:

\begin{equation}
\xi \propto \exp{[-1/\sqrt{(V/t)-(V/t)_c}]}.
\end{equation}
Rearranging terms and considering that, sufficiently close to $(V/t)_c$, $\xi$ can be replaced by the entire system size $L$,
we obtain that the critical interaction of the liquid-crystal phase transition satisfies the following scaling law:

\begin{equation}
(V/t)_c(L) \propto \dfrac{1}{\mathrm{ln}^2L}.
\end{equation}
In the inset of Fig.~\ref{fig:BKT}, we recover this scaling and we extrapolate with a linear fit the thermodynamic limit transition point to be $(V/t)_c \simeq 4.0$. This value represents a lower bound of the actual critical interaction strength.
In Fig.~\ref{fig:BKT} we also show, for completeness, the evolution of the superfluid stiffness $\rho_s(L)$ and of the quantity $S_L(\pi/2)/L$ for the largest system size numerically
available, as functions of interactions (here $S_L(\pi/2)$ is the size-dependent static structure factor value at the density peak). In the liquid regime,
it is known that $\lim_{L\rightarrow\infty}\rho_s(L) \neq 0$ and $\lim_{L\rightarrow\infty}S_L(\pi/2)/L = 0$,
whereas $\lim_{L\rightarrow\infty}\rho_s(L) = 0$ and $\lim_{L\rightarrow\infty}S_L(\pi/2)/L \neq 0$ in the crystalline phase: even if
this method is not sufficiently accurate to produce a quantitative (alternative) estimate of $(V/t)_c$, it establishes a range ($4<V/t<7$) in which
the result of our previous method undoubtedly lies, confirming the reliability of our analysis.

\vspace{0.5cm}

\newpage

\section{Generality of results for different densities}

\begin{figure}[h!]
\centerline{\includegraphics[width=0.95\columnwidth]{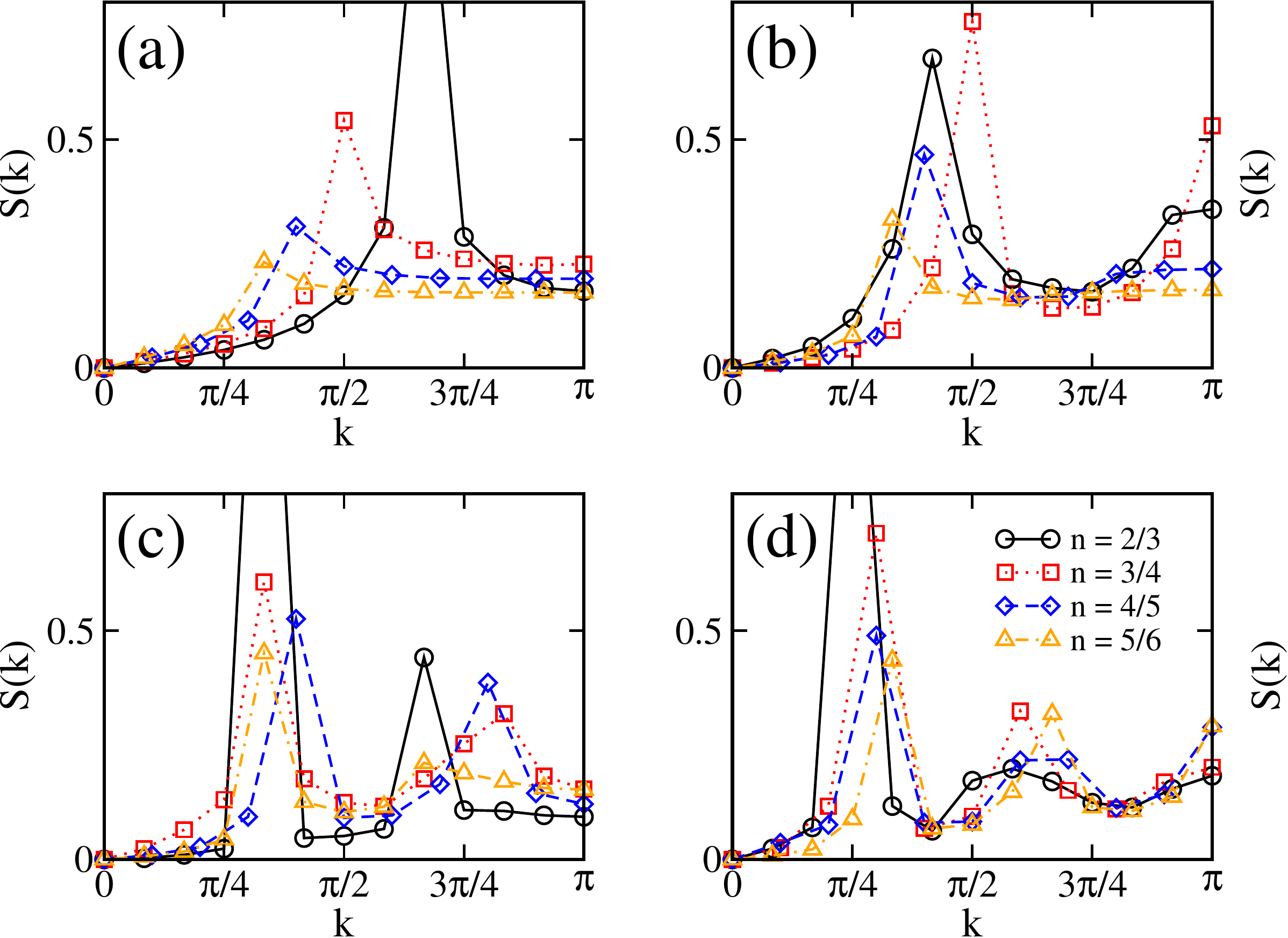}}
\caption{Static structure factor for $r_c=2$ [panel (a)], $r_c=3$ [panel (b)], $r_c=4$ [panel (c)] and $r_c=5$ [panel (d)] at densities $\overline{n}=2/3$ (black circles), $\overline{n}=3/4$
(red squares), $\overline{n}=4/5$ (blue diamonds), $\overline{n}=5/6$ (orange triangles). $V/t=4$ is kept fixed. }
\label{fig:fill}
\end{figure}

The numerical results we present in the main text for $\overline{n}=3/4$, are qualitatively reproducible also with other non-integer densities,
provided $r_c$ is sufficiently large to guarantee clustering (i.e., $r_c>r^{\star}$). \\
In Fig.~\ref{fig:fill}, we show $S(k)$ obtained from exact diagonalizations for different $\overline{n}$ and $r_c=1,2,3,4$ [panels (a),(b),(c) and (d), respectively].
As expected for, e.g., $\overline{n}=2/3$ (black circles), the peak shifts comparing the $r_c=r^{\star}=2$ case (TL liquid) with the $r_c =3$ case (cL liquid). 
Contrarily, for $\overline{n}=5/6$ (orange triangles), although the principal peak is sharpened increasing $r_c$, it remains at the same $k_c$. This shows that
density modes dominate in the $r_c$-parameter space investigated, where cL liquids would appear only for $r_c\geqslant6$.

\end{document}